\documentclass[prd,twocolumn,eqsecnum,showpacs,amsfonts,amssymb]{revtex4}

\usepackage{graphicx}
\input epsf

\usepackage{bm}

\newcommand{\beq}{\begin{equation}}
\newcommand{\eeq}{\end{equation}}
\newcommand{\beqs}{\begin{eqnarray}}
\newcommand{\eeqs}{\end{eqnarray}}

\newcommand{\gsim}{\mathrel{\raisebox{-
.6ex}{$\stackrel{\textstyle>}{\sim}$}}}

\begin{document}

\author{Shu-Chiuan Chang$^{a, b}$\footnote{(a): Permanent address; 
(b) Address on sabbatical} and Robert Shrock$^b$}

\title{Improved Lower Bounds on the Ground-State Entropy of the 
Antiferromagnetic Potts Model} 

\affiliation{(a) \ Department of Physics, National Cheng Kung University,
Tainan 70101, Taiwan} 

\affiliation{(b) \ C. N. Yang Institute for Theoretical Physics and
Department of Physics and Astronomy \\
Stony Brook University, Stony Brook, NY 11794, USA }

\begin{abstract}

We present generalized methods for calculating lower bounds on the ground-state
entropy per site, $S_0$, or equivalently, the ground-state degeneracy per site,
$W=e^{S_0/k_B}$, of the antiferromagnetic Potts model. We use these methods to
derive improved lower bounds on $W$ for several lattices.

\end{abstract}

\pacs{02.10.Ox,05.50.+q,64.60.De,75.10.Hk}

\maketitle

% =======================================================================

\section{Introduction}
\label{intro}

Nonzero ground-state entropy, $S_0 \ne 0$, is an important phenomenon 
in statistical mechanics.  An example of this is water ice, for which 
$S_0 = 0.82 \pm 0.05$ cal/(K-mole), i.e.,
$S_0/R = 0.41 \pm 0.03$ \cite{i35}-\cite{berg07}.  A model with $S_0 \ne 0$
is the $q$-state Potts antiferromagnet (AF) on a lattice
$\Lambda$ for sufficiently large $q$ \cite{wurev}.  This subject also has an
interesting connection with graph theory, since the partition function
of the $q$-state Potts antiferromagnet at temperature $T=0$ on a graph $G$
is
\beq
Z(G,q,T=0)_{PAF}=P(G,q) \ , 
\label{zp}
\eeq
where $P(G,q)$ is the chromatic polynomial of $G$, which is equal to the number
of ways of coloring the vertices of $G$ with $q$ colors subject to the
constraint that adjacent vertices must have different colors.  Such a color
assignment is called a proper vertex $q$-coloring of $G$. The minimum number of
colors required for a proper vertex $q$-coloring of the graph $G$ is called the
chromatic number of the graph, denoted $\chi(G)$.  We will focus here on
regular $N$-vertex lattice graphs $\Lambda_N$ and, in particular, on the
thermodynamic limit $N \to \infty$ (with appropriate boundary conditions),
which will be denoted simply as $\Lambda$.  In this limit, the ground-state
(i.e., zero-temperature) degeneracy per vertex (site) of the $q$-state Potts
antiferromagnet on $\Lambda$ is given by
\beq
W(\Lambda,q) = \lim_{N \to \infty} P(\Lambda_N,q)^{1/N} \ , 
\label{w}
\eeq
and the associated ground-state entropy per site is given by $S_0(\Lambda,q) =
k_B \, \ln W(\Lambda,q)$. It will be convenient to express our bounds on the
ground-state entropy per site in terms of its exponent, 
$e^{S_0(\Lambda,q)/k_B} = W(\Lambda,q)$.

In \cite{ww,w3} by S.-H. Tsai and one of us (RS), lower bounds on
$W(\Lambda,q)$ were derived for the triangular ($tri$), honeycomb ($hc$), $(4
\cdot 8^2)$, and $sq_d$ lattices. Here an Archimedean lattice $\Lambda$ is
defined as a uniform tiling of the plane with a set of regular polygons such
that all vertices are equivalent. Our notation for an Archimedean lattice
follows the standard mathematical format \cite{wn,gsbook}, namely $\Lambda =
(\prod_i p_i^{a_i})$, where the product is over the regular polygons $p_i$ that
are traversed in a circuit around a vertex and $a_i \ge 1$ refers to possible
contiguous repetitions of a given type of polygon in such a traversal.  The
$sq_d$ lattice is a nonplanar lattice formed from the square lattice by adding
edges (bonds) connecting the two sets of diagonal next-nearest-neighbor
vertices in each square.  In \cite{wn}, Shrock and Tsai derived corresponding
lower bounds on $W(\Lambda,q)$ for all Archimedean lattices $\Lambda$ and their
planar duals, using a coloring compatibility matrix (CCM) method employed
earlier by Biggs for the square ($sq$) lattice \cite{biggs77}, in combination
with the Perron-Frobenius theorem \cite{pf} and a theorem giving a lower bound
on the maximal eigenvalue of a symmetric non-negative matrix \cite{london}.

In this paper we introduce several generalizations of the method used in
\cite{ww}-\cite{biggs77} and apply these to derive improved lower bounds on
$W(\Lambda,q)$ for several lattices $\Lambda$.  Refs. \cite{biggs77} and
\cite{ww,w3} also used CCM methods to derive upper bounds on
$W(\Lambda,q)$. However, it was shown in \cite{ww,w3} that, while the upper
bounds were moderately restrictive, the lower bounds were very close to the
actual values of $W(\Lambda,q)$.  Therefore, as in \cite{wn}, we focus here on
the lower bounds on $W(\Lambda,q)$.

This paper is organized as follows. In Section \ref{method} we explain the
basic coloring compatibility matrix method. In Section \ref{generalized_method}
we discuss our generalizations of this method.  In Sections \ref{sq}-\ref{hc}
we apply our generalized methods to derive new and more restrictive lower
bounds on $W(\Lambda,q)$ for the square, triangular, and honeycomb lattices.
In Sections \ref{488} and \ref{kag} we present corresponding results for two
heteropolygonal Archimedean lattices, namely, the $(4 \cdot 8^2)$ and $(3 \cdot
6 \cdot 3 \cdot 6)$ (i.e., kagom\'e) lattices.  In Section \ref{sqd} we report
results for the $sq_d$ lattice.  In Section \ref{small_y_series} we compare the
large-$q$ Taylor series expansions of our lower bounds for the various lattices
with the large-$q$ series expansions of the actual $W$ functions for these
respective lattices.  Our conclusions are given in Section
\ref{conclusions}. We list some results on $r$-partite lattices in Appendix
\ref{rpartite}, the lower bounds on $W(\Lambda,q)$ for Archimedean lattices
$\Lambda$ from \cite{ww}-\cite{biggs77} in Appendix \ref{wnbounds}, and some
higher-degree algebraic equations that are used in the text in Appendix
\ref{higherdegree}.

% =========================================================================

\section{Basic Calculational Method}
\label{method}

  In this section we explain the basic calculational method used in
\cite{ww}-\cite{biggs77} to derive lower bounds on $W(\Lambda,q)$.  In the next
section we generalize this method in several ways. We consider a sequence of
(regular) lattices of type $\Lambda$ of length $L_x=n$ vertices in the
longitudinal direction and width $L_y=m$ vertices in the transverse direction.
In the thermodynamic limit $n \to \infty$, $m \to \infty$ with the
aspect ratio $m/n$ finite, the boundary conditions do not affect
$W(\Lambda,q)$. It will be convenient to take periodic boundary conditions
(PBCs) in both directions. If a lattice $\Lambda$ is $r$-partite, then $m$ and
$n$ are chosen so as to maintain this property. 

The construction of the coloring compatibility matrix $T$ begins by considering
an $n$-vertex path ${\cal P}_n$ in the longitudinal direction on
$\Lambda$. The number of proper vertex $q$-colorings of ${\cal P}_n$ is the
chromatic polynomial $P({\cal P}_n,q)$. Now focus on two adjacent parallel
paths, ${\cal P}_n$ and ${\cal P}'_n$.  Define compatible proper
$q$-colorings of the vertices of these adjacent paths as proper $q$-colorings
such that no two adjacent vertices on ${\cal P}_n$ and ${\cal P}'_n$
have the same color.  One can then associate with this pair of adjacent paths
an ${\cal N} \times {\cal N}$ dimensional symmetric matrix $T$, where ${\cal N}
= P({\cal P}_n,q)=P({\cal P}'_n,q)$, with entries $T_{{\cal
P}_n,{\cal P}'_n} = T_{{\cal P}'_n,{\cal P}_n} =$ 1 or 0 if the
proper $q$-colorings of ${\cal P}_n$ and ${\cal P}'_n$ are or are not
compatible, respectively.  This matrix is thus defined in the space of allowed
color configurations for these adjacent paths.

It follows that, for fixed $m$ and $n$, 
\beq
P(\Lambda_{m \times n},q) = {\rm Tr}(T^m) \ .
\label{plambda}
\eeq
For a given $n$, since $T$ is a nonnegative matrix, one can apply the
Perron-Frobenius theorem \cite{pf} to conclude that $T$ has a real positive
maximal eigenvalue $\lambda_{max}$.  Hence, for fixed $n$,
\beq
\lim_{m \to \infty} Tr(T^m)^{\frac{1}{mn}} = 
(\lambda_{max})^{\frac{1}{n}} \ . 
\label{trlim}
\eeq
Therefore, taking the $n \to \infty$ limit, 
\beq
W(\Lambda,q) = \lim_{n \to \infty} (\lambda_{max})^{\frac{1}{n}} \ . 
\label{wlim}
\eeq

Let us denote the column sum 
\beq
\kappa_j(T) = \sum_{i=1}^{\cal N} T_{ij} \ , 
\label{kappa}
\eeq
which is equal to the row sum, 
\beq
\rho_j(T)=\sum_{i=1}^{\cal N}T_{ji} \ , 
\label{rho}
\eeq
(since $T^T=T$) and the sum of all entries of $T$ as 
\beq
S(T) = \sum_{i,j=1}^{\cal N} T_{ij} \ . 
\label{st}
\eeq
Note that $S(T)/{\cal N}$ is the average row sum (equal to the average column
sum).  

For a general nonnegative ${\cal N} \times {\cal N}$ matrix $A$,
\cite{pf}, one has 
\beq
\min\{\kappa_j(A) \} \le \lambda_{max}(A) \le \max\{\kappa_j(A) \}
\label{kappa_inequality}
\eeq
and
\beq
\min\{\rho_j(A) \} \le \lambda_{max}(A) \le \max\{\rho_j(A) \}
\label{rho_inequality}
\eeq
for $j=1,...,{\cal N}$. Since $T^T=T$, these are equivalent here.  One also
has the following more restrictive one-parameter family of lower bounds
depending on the parameter $k$, for a symmetric nonnegative matrix $T$
\cite{london}:
\beq
\biggl [\frac{S(T^k)}{\cal N} \biggr ]^{1/k} \le \lambda_{max}(T) \ . 
\label{lbound}
\eeq

Refs. \cite{ww}-\cite{biggs77} derived lower and upper bounds on $W(\Lambda,q)$
using the $k=1$ special case of (\ref{lbound}).  We will denote a generic lower
bound on $W(\Lambda,q)$ with the subscript $\ell$, as $W(\Lambda,q)_\ell$. We
will distinguish specific lower bounds that we obtain with the additional
subscripts $b$ and $k$, as explained below.  The lower bounds obtained in
\cite{ww}-\cite{biggs77} were for $b=1$ and $k=1$.  Refs. \cite{ww}-\cite{wn}
studied how close the upper and lower bounds obtained on $W(\Lambda,q)$ were to
the actual values of $W(\Lambda,q)$ for a number of lattices, where the latter
were determined mainly from Monte Carlo calculations, augmented by large-$q$
series expansions together with a few exact results.  It was found that for a
given lattice $\Lambda$, as $q$ increases beyond the region of 
$\chi(\Lambda)$, the lower bounds rapidly approach very close to the actual
value of $W(\Lambda,q)$. 

We next introduce some notation that will be used below for reduced functions
obtained from $W(\Lambda,q)$ which will be analyzed in the large-$q$ limit.
This large-$q$ limit is the natural one to consider for chromatic polynomials,
since the constraint in a proper $q$-coloring of the vertices of a graph,
namely that no two adjacent vertices have the same color, becomes progressively
less restrictive as the number of colors increases to large values. The
chromatic polynomial of an arbitrary $N$-vertex graph $G$ is a polynomial of
degree $N$, and consequently, $W(\Lambda,q) \sim q$ as $q \to \infty$. In order
to deal with a finite quantity in the $q \to \infty$ limit, one therefore
considers the reduced ($r$) function $W_r(\Lambda,q) = W(\Lambda,q)/q$.  A
variable equivalent to $1/q$ that is convenient to use for a large-$q$ series
expansion of $W_r(\Lambda,q)$ is $y = 1/(q-1)$.  These large-$q$ (i.e.,
small-$y$) series expansions are normally given for the function
\beq
\overline W(\Lambda,y) = \frac{W_r(\Lambda,q)}{(1-q^{-1})^{\Delta_\Lambda/2}} 
= \frac{W(\Lambda,q)}{q(1-q^{-1})^{\Delta_\Lambda/2}} \ ,
\label{wbar}
\eeq
where $\Delta_\Lambda$ is the lattice coordination number of the lattice
$\Lambda$ (i.e., the degree of the vertices of $\Lambda$).  In terms
of the expansion variable $y$, these series thus have the form
$\overline W(\Lambda,y)=1+\sum_{k=1}^\infty w_{\Lambda,k} y^k$. 
Analogously, for the expansion of our lower bound, we define the reduced
lower bound function $\overline W(\Lambda,y)_\ell$ as
\beq
\overline W(\Lambda,y)_\ell = \frac{W(\Lambda,q)_\ell}
{q(1-q^{-1})^{\Delta_\Lambda/2}}
\ . 
\label{wlbseriesdef}
\eeq

Before proceeding, we note a subtlety in the definition of $W(\Lambda,q)$.  As
pointed out in \cite{w}, the formal eq. (\ref{w}) is not, in general, adequate
to define $W(\Lambda,q)$ because of a noncommutativity of limits
\beq 
\lim_{N \to \infty} \lim_{q \to q_s} P(\Lambda_N,q)^{1/N} \ne \lim_{q
\to q_s} \lim_{N \to \infty} P(\Lambda_N,q)^{1/N}
\label{wnoncomm}
\eeq
at certain special points $q_s$.  We denote the definitions based on the first
and second orders of limits in (\ref{wnoncomm}) as $W(\Lambda,q)_{D_{Nq}}$ and
$W(\Lambda,q)_{D_{qN}}$, respectively.  This noncommutativity can occur for $q
< q_c(\Lambda)$, where $q_c(\Lambda)$ denotes the maximal (finite) real value
of $q$ where $W(\Lambda,q)$ is nonanalytic \cite{w}.  These values include
$q_c(sq)=3$, $q_c(tri)=4$, and the formal value $q_c(hc)=(3+\sqrt{5})/2 =
2.618...$ \cite{wurev,w} for the square, triangular, and honeycomb lattices.
As explained in \cite{w}, the underlying reason for the noncommutativity is
that as $q$ decreases from large values, there is a change in the analytic
expression for $W(\Lambda,q)$ as $q$ decreases through the value
$q_c(\Lambda)$. We do not have to deal with this complication here because
elementary results yield exact values of $W(sq,2)$, $W(hc,2)$, $W((4 \cdot
8^2),2)$, and $W(tri,3)$
(see Eqs. (\ref{pgrpartite}) and (\ref{wrpartite})), namely
\beq
W(sq,2)=W(hc,2)=W((4 \cdot 8^2),2) = 1, \quad W(tri,3)=1 \ . 
\label{wrpartite_lambdas}
\eeq
Hence, our lower bounds are not needed at the
respective values $q=2$ for the square, honeycomb, and $(4 \cdot 8^2)$ lattices
or for $q=3$ on the triangular lattice, and we therefore focus on their
application to $q \ge 3$ for $\Lambda=sq, \ hc, \ (4 \cdot 8^2)$ and to $q \ge
4$ for $\Lambda=tri$, and similarly for other lattices. 

% ======================================================================

\section{Generalized Coloring Compatibility Matrix Method}
\label{generalized_method}

% ======================================================================

\subsection{Coloring Compatibility Matrix Joining Adjacent Strips of Width $b$}
\label{bbk1}

The lower bounds on $W(\Lambda,q)$ derived in \cite{ww}-\cite{biggs77} for
various lattices $\Lambda$ used Eq. (\ref{lbound}) with $T$ being a coloring
compatibility matrix joining adjacent paths and with $k=1$. Here we generalize
this method in several ways. Our first generalization is to use a coloring
compatibility matrix that joins adjacent strips of width $b \ge 2$ vertices,
rather than adjacent one-dimensional ($b=1$) paths.  For simplicity, we explain
this for the square lattice; similar discussions apply for other lattices. We
define the matrix $T$ to enumerate compatible colorings of a strip of
transverse width $b$ vertices and an adjacent parallel strip of width $b$ and
arbitrary length $L_x$ vertices, with cyclic boundary conditions. (Here, by
cyclic boundary conditions for a given strip, we mean in the $x$, i.e.,
longitudinal, direction along this strip). The condition that these strips are
adjacent is equivalent to the statement that they share a common set of edges.
Thus, this CCM is an ${\cal N} \times {\cal N}$ matrix, where ${\cal N}$ is the
chromatic polynomial for the cyclic strip of width $b$ vertices and arbitrary
length $L_x$, with cyclic boundary conditions. For this CCM, the sum of
elements $S(T)$ is equal to the chromatic polynomial of a strip of width
$L_y=2b-1$ vertices and arbitrary length $L_x$ vertices with cyclic boundary
conditions. These chromatic polynomials of lattice strips of a fixed width
$L_y$ and arbitrarily great length $L_x$ with periodic boundary conditions in
the longitudinal direction and free boundary conditions in the transverse
direction have the form
\beq
P(\Lambda,L_y \times L_x,cycl.,q) = \sum_{d=0}^{L_y} c^{(d)}
\sum_{j=1}^{n_P(L_y,d)} (\lambda_{sq,L_y,d,j})^{L_x} 
\label{pformcyc}
\eeq
with 
\beq
c^{(d)} = \sum_{j=0}^d (-1)^j \, {2d-j \choose j} \, q^{d-j}  \ , 
\label{cd}
\eeq
where ${a \choose b} = a!/[b!(a-b)!]$ is the binomial coefficient.  
For a table
of the $n_P(L_y,d)$, see \cite{cf}. Because of the limits (\ref{trlim}) and
(\ref{wlim}), only the largest $\lambda_{\Lambda,L_y,d,j}$ enters in the lower
bound (\ref{lbound}) in the thermodynamic limit.  As specific studies such as
\cite{wcy}-\cite{k} showed, the dominant $\lambda$ for the values of $q$ of
relevance here is $\lambda_{\Lambda,L_y,0,1}$.

Applying this generalization of the coloring compatibility matrix in
combination with the $k=1$ case of (\ref{lbound}), we derive the new 
lower bound for $b \ge 2$: 
\beq
W(\Lambda,q) \ge W(\Lambda,q)_{\ell;b,1} \ , 
\label{wlbound_bbk1}
\eeq
where
\beq
W(\Lambda,q)_{\ell;b,1} = \Bigg [\frac{\lambda_{\Lambda,2b-1,0,1}}
{\lambda_{\Lambda,b,0,1}} \Bigg ]^{\frac{1}{b-1}} \ . 
\label{wlbound_value_bbk1}
\eeq
The final subscript, 1, in $W(\Lambda,q)_{\ell;b,1}$ in 
(\ref{wlbound_bbk1}) and (\ref{wlbound_value_bbk1}) 
is the value of $k$.

The corresponding lower bound for $\overline W(\Lambda,y)$ is
\beq
\overline W(\Lambda,y) \ge \overline W(\Lambda,y)_{\ell;b,1} \ , 
\label{wylbound_bbk1}
\eeq
where, in accordance with Eq. (\ref{wbar}), 
\beq
\overline W(\Lambda,y)_{\ell;b,1} = 
\frac{W(\Lambda,q)_{\ell;b,1}}{q(1-q^{-1})^{\Delta_\Lambda/2}} \ , 
\label{wylbound_value_bbk1}
\eeq
with $\Delta_\Lambda$ being the coordination number of the lattice $\Lambda$,
as before.  The inequality (\ref{wlbound_bbk1}) with (\ref{wlbound_value_bbk1})
is actually an infinite family of lower bounds depending on the strip width
$b=1,2,...$, and similarly with (\ref{wylbound_bbk1}) and
(\ref{wylbound_value_bbk1}).  This is one of our two major results, which we
will proceed to apply to a number of different lattices.  The special case
$b=1$ was previously used in \cite{biggs77} and \cite{ww}-\cite{wn} to derive
lower bounds which we denote here as $W(\Lambda,q)_{\ell;1,1}$ and
correspondingly $\overline W(\Lambda,y)_{\ell;1,1}$. Our generalization in this
subsection is to $b \ge 2$ with $k=1$.

% ====================================================================

\subsection{Coloring Compatibility Matrix Acting $k$ Times Joining Paths of
  Width $b=1$ }
\label{kkb2}

Our second generalization is to use a coloring compatibility matrix method 
that involves paths (i.e., one-dimensional strips, with $b=1$) on $\Lambda$
that are separated by $k$ edges, where $k \ge 2$, rather than the situation
with $b=1$ and $k=1$ considered in \cite{ww}-\cite{wn}, where the paths were
adjacent.  This means using the coloring compatibility matrix $T$ defined as
connecting adjacent paths, and having it operate $k$ times, with $k \ge 2$.
Hence, ${\cal N}=P(C_{L_x},q)$ and $S(T^k)$ is the chromatic polynomial of a
strip of width $L_y=k+1$ vertices and arbitrary length $L_x$ vertices with
cyclic boundary conditions.  Again, only the dominant
$\lambda_{\Lambda,L_y,d,j}$ terms enter in (\ref{lbound}) in the thermodynamic
limit. Using this method in combination with (\ref{lbound}), we derive the
lower bound
\beq
W(\Lambda,q) \ge W(\Lambda,q)_{\ell;1,k} \ , 
\label{wlbound_b1kk}
\eeq
where 
\beq
W(\Lambda,q)_{\ell;1,k} = 
\Bigg [ \frac{\lambda_{\Lambda,k+1,0,1}}{\lambda_{\Lambda,1,0,1}} 
\Bigg ]^{\frac{1}{k}} \ . 
\label{wlbound_b1kk_value}
\eeq
In (\ref{wlbound_b1kk}) and (\ref{wlbound_b1kk_value}), the first subscript
after $\ell;$ is $b=1$.

An important theorem extending the result (\ref{lbound}) is that for a
symmetric nonnegative matrix $T$ \cite{merikoski84}, 
\beq
\biggl [\frac{S(T^k)}{\cal N} \biggr ]^{1/k} \quad {\rm is \ an \ increasing \
function \ of} \ k \ .
\label{merikoski}
\eeq
It follows that, for the physical range of $q$ of relevance for our application
to a lattice $\Lambda$, 
\beq
W(\Lambda,q)_{\ell;1,k} \quad {\rm is \ an \ increasing \ function \ of} \ 
k \ .
\label{wlb_b1kk_increasing}
\eeq

The corresponding lower bound for $\overline W(\Lambda,y)$ is
\beq
\overline W(\Lambda,y) 
\ge \overline W(\Lambda,y)_{\ell;1,k} \ , 
\label{wylbound_b1kk}
\eeq
where, in accordance with Eq. (\ref{wbar}), 
\beq
\overline W(\Lambda,y)_{\ell;1,k} = 
\frac{W(\Lambda,q)_{\ell;1,k}}{q(1-q^{-1})^{\Delta_\Lambda/2}} \ . 
\label{wylbound_value_b1kk}
\eeq
Again, the inequality (\ref{wlbound_b1kk}) with (\ref{wlbound_b1kk_value}) is
actually a one-parameter family of lower bounds depending on the parameter
$k=1,2,...$, and similarly with (\ref{wylbound_b1kk}) and
(\ref{wylbound_value_b1kk}).  This is the second of our major results.  The
special case $k=1$ (with $b=1$) was previously used in
\cite{ww}-\cite{biggs77}; the generalization presented in this subsection is to
$k \ge 2$ with $b=1$. We have also carried out further generalizations of lower
bounds on $W(\Lambda,q)$ with both $b \ge 2$ and $k \ge 2$.  These are more
complicated and will be presented elsewhere.

% =========================================================================

\subsection{Measures of Improvement of Bounds}

For a lattice $\Lambda$ and a given $q$, we define the ratio of a lower bound
$W(\Lambda,q)_{\ell;b,k}$ to the actual value of $W(\Lambda,q)$ as
\beq
R_{\Lambda,q;\ell;b,k} \equiv \frac{W(\Lambda,q)_{\ell;b,k}}{W(\Lambda,q)} \ .
\label{rbk}
\eeq
This ratio is useful as a measure of how close a particular lower bound
$W(\Lambda,q)_{\ell;b,k}$ is to the actual value of the ground-state degeneracy
per vertex, $W(\Lambda,q)$.  For most lattices and values of $q$, the value of
$W(\Lambda,q)$ is not known exactly, but rather is determined for moderate
values of $q$ by Monte Carlo simulations, as discussed in \cite{ww,w3} and, for
larger values of $q$, by large-$q$ series expansions \cite{kimenting}.  Special
cases of $\Lambda$ and $q$ for which exact results are known will be noted
below.

An important property of our new lower bounds is that, for a given lattice
$\Lambda$, they are larger than and hence more restrictive than the bounds
$W(\Lambda,q)_{\ell;1,1}$ derived in \cite{ww}-\cite{biggs77}.  Since the
lower bounds $W(\Lambda,q)_{\ell;1,1}$ were very close to the actual values 
of $W(\Lambda,q)$ for all but the lowest values of $q$, our improved lower 
bounds are even closer to these actual values.  For the same reason, our new
lower bounds yield the greatest fractional improvement for low to moderate
values of $q$ and are only slightly greater than $W(\Lambda,q)_{\ell;1,1}$ for
larger values of $q$.  This will be evident in our explicit results. 
For our present discussion, we take $T$ to be the matrix that acts $k$
times mapping a strip of width $b$ to an adjacent strip of width $b$ on
$\Lambda$.  Then the theorem (\ref{merikoski}) and its corollary
(\ref{wlb_b1kk_increasing}) imply that, for fixed $b$, the ratio of our lower
bound $W(\Lambda,q)_{\ell;b,k}$ to the actual value $W(\Lambda,q)$ is an
increasing function of $k$, i.e.,
\beq
R_{\Lambda,q;\ell;b,k} \quad {\rm is \ an \ increasing \ function \ of} \ k \ .
\label{rb1kk_k_increasing}
\eeq
That is, as $k$ increases, the lower bound $W(\Lambda,q)_{\ell;b,k}$
becomes more restrictive.  From our analysis, we also find that for fixed
$k=1$ and $b \ge 2$,
\beq
R_{\Lambda,q;\ell;b,1} \quad {\rm is \ an \ increasing \ function \ of} \ b \ .
\label{rbbk1_b_increasing}
\eeq

For a given $\Lambda$ and $q$, it is also of interest to compare the various
lower bounds with each other.  For this purpose, we define the ratio
\beq
R_{\Lambda,q;(b,k)/(b',k')} \equiv \frac{W(\Lambda,q)_{\ell;b,k}}
{W(\Lambda,q)_{\ell;b',k'}} \ .
\label{rbkbpkp}
\eeq
By the same argument, theorem (\ref{merikoski}) and its corollary
(\ref{wlb_b1kk_increasing}) imply that for a given lattice $\Lambda$, our
new lower bounds $W(\Lambda,q)_{\ell;1,k}$ improve on the bound
$W(\Lambda,q)_{\ell;1,1}$ derived in \cite{ww}-\cite{biggs77}:
$W(\Lambda,q)_{\ell;1,k} \ge W(\Lambda,q)_{\ell;1,1}$, i.e.,
\beq
R_{\Lambda,q;(1,k)/(1,1)} \ge 1 \quad {\rm for} \ k \ge 2 \ . 
\label{rb1kk_over_b1k1}
\eeq
We observe also that
\beq
R_{\Lambda,q;(b,1)/(1,1)} \ge 1 \quad {\rm for} \ b \ge 2 \ . 
\label{rbbk1_over_b1k1}
\eeq
As will be evident from our explicit results, for the range of $q$ that we
consider, these inequalities are realized as strict inequalities.  As noted
above, since the latter lower bounds $W(\Lambda,q)_{\ell;1,1}$ are very close
to the actual values of $W(\Lambda,q)$, even for $q$ only moderately above
$\chi(\Lambda)$, as shown in Table I of \cite{ww} and Tables I-III of
\cite{w3}, our new bounds are even closer to these actual values of
$W(\Lambda,q)$. In all cases, we find that the ratios approach unity rapidly 
in the limit $q \to \infty$.

A major result of Ref. \cite{wn} was the derivation of general formulas for the
lower bound $W(\Lambda,q)_{\ell;1,1}$ and $\overline W(\Lambda,y)_{\ell;1,1}$
for all Archimedean lattices and their (planar) duals (Eqs. (4.11), (4.13),
(5.1), and (5.2) in \cite{wn}).  As will be evident below, aside from the basic
theorems, our new lower bounds $W(\Lambda,q)_{\ell;b,k}$ with $b \ge 2$ and/or
$k \ge 2$ do not have such simple general formulas.  However, as noted, they
do provide a useful improvement on the earlier $W(\Lambda,q)_{\ell;1,1}$ lower
bounds, especially for $q$ values not too much larger than $\chi(\Lambda)$. 

% ========================================================================

\section{Square Lattice}
\label{sq}

As noted above, since the value $W(sq,2)=1$ is known
exactly by elementary methods, we focus on the application of our new
lower bounds to the range $q \ge 3$. We first recall the result for the case
$b=1$, $k=1$. With $T$ being the coloring matrix connecting adjacent rows or
columns of a square lattice, and with the application of the $k=1$ special case
of the theorem (\ref{lbound}), one has
\beq
W(sq,q) \ge W(sq,q)_{\ell;1,1} \ , 
\label{wsqlbound_b1k1}
\eeq
where \cite{biggs77} 
\beq
W(sq,q)_{\ell;1,1} =\frac{q^2-3q+3}{q-1} \ . 
\label{wsqlbound_b1k1_value}
\eeq
In terms of $\overline W(sq,y)$, given by (\ref{wbar}) with $\Lambda=sq$ and
$\Delta=4$, the lower bound is the $b=1$ case of 
(\ref{wylbound_bbk1}) with $\Lambda=sq$, namely 
\beq
\overline W(sq,y)_{\ell;1,1} =1+y^3 \ , 
\label{wysqlbound_b1k1_value}
\eeq
as listed in Table III of \cite{wn}. 

% ==================================================================

\subsection{ CCM Method with $b=2, \ 3$ and $k=1$} 

We first use our generalized method with the coloring compatibility matrix
relating the allowed colorings of a width $b=2$ cyclic ladder strip of the
square lattice to those of the adjacent $b=2$ strip. For this, we need the
dominant term in the chromatic polynomial for the square-lattice strip of width
$2b-1=3$ for the relevant range of $q \ge 3$.  This chromatic polynomial was
calculated in \cite{wcy}, and the dominant term is $\lambda_{sq,3,0,1}$, namely
\beqs
& & \lambda_{sq,3,0,1} = \frac{1}{2} \, \bigg [ (q-2)(q^2-3q+5) \cr\cr
& + & \Big [ (q^2-5q+7)(q^4-5q^3+11q^2-12q+8) \Big ]^{1/2} \ \bigg ] \ .
\cr\cr
& & 
\label{lamsqly3}
\eeqs
This term is also the dominant $\lambda$ in the chromatic polynomial for the
strip of the square lattice with transverse width $L_y=3$ vertices and arbitary
length, with free longitudinal and transverse boundary conditions \cite{strip}.
Our lower bound with $b=2$ (and $k=1$) then reads
\beq
W(sq,q) \ge W(sq,q)_{\ell;2,1} \ , 
\label{wsqlbound_b2k1}
\eeq
where (with $\lambda_{sq,2,0,1}=q^2-3q+3$) \cite{wsqlbound_b2k1_analytic}
\begin{widetext}
\beq
W(sq,q)_{\ell;2,1} = \frac{\lambda_{sq,3,0,1}}{\lambda_{sq,2,0,1}} 
= \frac{(q-2)(q^2-3q+5)
+ \Big [ (q^2-5q+7)(q^4-5q^3+11q^2-12q+8) \Big ]^{1/2}}{2(q^2-3q+3)}  \ . 
\label{wsqlbound_b2k1_value}
\eeq
\end{widetext}

Using the analytic results (\ref{wsqlbound_b1k1_value}) and
(\ref{wsqlbound_b2k1_value}), we have proved the following inequality 
(for $q \ge 3$):
\beq
W(sq,q)_{\ell;2,1} \ge W(sq,q)_{\ell;1,1} 
\label{wsqlbound_b2k1_ge_wsqlbound_b1k1}
\eeq
In terms of the ratio $R_{sq,q;(2,1)/(1,1)}$, 
\beq
R_{sq,q;(2,1)/(1,1)} \ge 1
\label{rsqlbound_b2k1_over_b1k1}
\eeq
The inequality (\ref{rsqlbound_b2k1_over_b1k1}) means that our new lower bound,
(\ref{wsqlbound_b2k1}), is more stringent than the previous lower bound
(\ref{wsqlbound_b1k1}) obtained with the CCM method with $b=1$ and $k=1$.

Ref. \cite{ww} showed that as $q$ increases, $R_{sq,q;1,1}$ rapidly approaches
extremely close to unity. For example, for $q=4, \ 5, \ 6$, $R_{sq,q;1,1}$ is
equal to 0.9984, 0.9997, and 0.9999 (see Table I in \cite{ww}), respectively,
and it increases monotonically with larger $q$.  Our improved lower bound
(\ref{wsqlbound_b2k1}) on $W(sq,q)$ is therefore even closer to the respective
actual values of $W(sq,q)$. As will be discussed below, this is also true of
our other new lower bounds using $b=1$ and $k \ge 2$.  We note that if one were
formally to extend the range of applicability of
(\ref{wsqlbound_b2k1_ge_wsqlbound_b1k1}) down to $q=2$, it would be realized as
an equality, and if one were to extend the range of applicability of
(\ref{rsqlbound_b2k1_over_b1k1}) to $2 \le q \le \infty$, it would be realized
as an equality at $q=2$ and in the limit $q \to \infty$.

For the previous lower bound $W(sq,q)_{\ell;1,1}$, the largest deviation from
the actual value occurs at $q=3$.  It happens that for $q=3$, $W(sq,3)$ is
known exactly \cite{lenard}:
\beq
W(sq,3) = \frac{8}{3^{3/2}} = 1.5396007...
\label{wsq_q3}
\eeq
For the old bound, 
\beq
W(sq,3)_{\ell;1,1}{}\Big |_{q=3} = \frac{3}{2} \ , 
\label{wsqlbound_b1k1_value_q3}
\eeq
so that 
\beq
\frac{W(sq,3)_{\ell;1,1}{}\Big |_{q=3}}{W(sq,3)} = \frac{3^{5/2}}{16} =
0.974279
\label{wsqlbound_b1k1_value_q3_over_exactvalue}
\eeq
to the indicated floating point accuracy. As guaranteed by the general
inequality (\ref{rsqlbound_b2k1_over_b1k1}), our lower
bound (\ref{wsqlbound_b2k1}) with (\ref{wsqlbound_b2k1_value}) improves on
this. For $q=3$, we have
\beq
W(sq,3)_{\ell;2,1}{}\Big |_{q=3} = \frac{5+\sqrt{17}}{6} 
= 1.5205176..., 
\label{wsqlbound_b2k1_value_q3}
\eeq
so that 
\beq
R_{sq,3;\ell;2,1} \equiv \frac{W(sq,3)_{\ell;2,1}{}\Big |_{q=3}}{W(sq,3)} = 
\frac{\sqrt{3} \, (5+\sqrt{17} \ )}{16} = 0.987605...
\label{wsqlbound_b2k1_value_q3_over_exactvalue}
\eeq
We show these ratios $R_{sq,3;\ell;1,1}$ and $R_{sq,3;\ell;2,1}$ in 
Table \ref{rsqlbound_bk_q3_table}. 

In terms of the function $\overline W(sq,y)$, our lower bound 
(\ref{wsqlbound_b2k1}) reads 
\beq
\overline W(sq,y) \ge \overline W(sq,y)_{\ell;2,1} \ , 
\label{wysqlbound_b2k1}
\eeq
where
\begin{widetext}
\beq
\overline W(sq,y)_{\ell;2,1} = \frac{(1+y)\bigg [ (1-y)(1-y+3y^2) + 
\Big [(1-3y+3y^2)(1-y+2y^2-y^3+3y^4) \Big ]^{1/2} \ \bigg ]}
{2(1-y+y^2)} \ . 
\label{wysqlbound_b2k1_value}
\eeq
We have also calculated the lower bound $W(sq,q)_{\ell;b,1}$ for $b=3$, and we
list the ratio $R_{sq,3;\ell;3,1}$ in Table \ref{rsqlbound_bk_q3_table}. 

% =======================================================================

\subsection{ CCM Method with $b=1$ and $2 \le k \le 5$} 

Next, we apply our second generalized method to the square lattice.  For $k=2$,
our lower bound obtained using this method is (\ref{wlbound_b1kk}) with 
(\ref{wlbound_b1kk_value}), namely 
\beq
W(sq,q) \ge W(sq,q)_{\ell;1,2} \ , 
\label{wsqlbound_b1k2}
\eeq
where
\beq
W(sq,q)_{\ell;1,2} = \bigg [ 
\frac{\lambda_{sq,3,0,1}}{\lambda_{C,0,1}} \bigg ]^{1/2} =
\Bigg [ \frac{(q-2)(q^2-3q+5) 
+ \Big [ (q^2-5q+7)(q^4-5q^3+11q^2-12q+8) \Big ]^{1/2}}{2(q-1)} \Bigg ]^{1/2}
\ . 
\label{wsqlbound_b1k2_value}
\eeq
The corresponding lower bound on $\overline W(sq,y)$ is 
\beq
\overline W(sq,y) \ge \overline W(sq,y)_{\ell;1,2} \ , 
\label{wysqlbound_b1k2}
\eeq
where 
\beq
\overline W(sq,y)_{\ell;1,2} = \frac{1}{\sqrt{2}} \, (1+y) 
\Bigg [ (1-y)(1-y+3y^2) + \Big [ (1-3y+3y^2)(1-y+2y^2-y^3+3y^4) \Big ]^{1/2} \ 
\Bigg ]^{1/2} \ . 
\label{wysqlbound_b1k2_value}
\eeq
\end{widetext}

For $b=1$ and $k=3$, we need the dominant $\lambda$ in the chromatic polynomial
for the cyclic square-lattice strip of width $L_y=4$ vertices and arbitrary
length $L_x$, namely $\lambda_{sq,4,0,1}$, which was calculated in \cite{s4}
(and is the same as the dominant $\lambda$ in the chromatic polynomial of the
free square-lattice strip of width $L_y=4$ \cite{strip}).  This term
$\lambda_{sq,4,0,1}$ is the largest (real) root of the cubic equation
(\ref{eqsq4}) in Appendix \ref{higherdegree}.  Our bound is then $W(sq,q) \ge
W(sq,q)_{\ell;1,3}$, where
\beq
W(sq,q)_{\ell;1,3} = \bigg [ \frac{\lambda_{sq,4,0,1}}{q-1} \bigg ]^{1/3} \ . 
\label{sqlbound_b1k3_value}
\eeq
In a similar manner, for $b=1$ and $k=4$, we have obtained the bound 
$W(sq,q) \ge W(sq,q)_{\ell;1,4}$, where 
\beq
W(sq,q)_{\ell;1,4} = \bigg [ \frac{\lambda_{sq,5,0,1}}{q-1} \bigg ]^{1/4} \ . 
\label{sqlbound_b1k4_value}
\eeq
and $\lambda_{sq,5,0,1}$ is the largest (real) root of an 
algebraic equation of degree 7. 

As a special case of our general result
(\ref{wlb_b1kk_increasing}), we have
\beq
R_{sq,q;(1,4)/(1,1)} \ge R_{sq,q;(1,3)/(1,1)} \ge  R_{sq,q;(1,2)/(1,1)} 
\ge 1 \ .
\label{r1211}
\eeq
In the range $q \ge 3$ under consideration here, we find that each 
$\ge$ is realized as $>$, i.e., a strict inequality. 

It is also of interest to compare our various lower bounds 
$W(sq,q)_{\ell;b,1}$ and $W(sq,q)_{\ell;1,k}$ with each other. For the first
two above the old case $b=1$, $k=1$, we find 
\beq
R_{sq,q;(2,1)/(1,2)} > 1 \ . 
\label{rsqlbound_b2k1_over_b1k2}
\eeq
That is, our lower bound with $(b,k)=(2,1)$ is larger, and hence more
restrictive, than our lower bound with $(b,k)=(1,2)$.  In the limit $q \to
\infty$, the ratio (\ref{rsqlbound_b2k1_over_b1k2}) approaches 1.  

% ==========================================================================

\subsection{Plots} 

\begin{figure}
  \begin{center}
    \includegraphics[height=8cm,width=6cm]{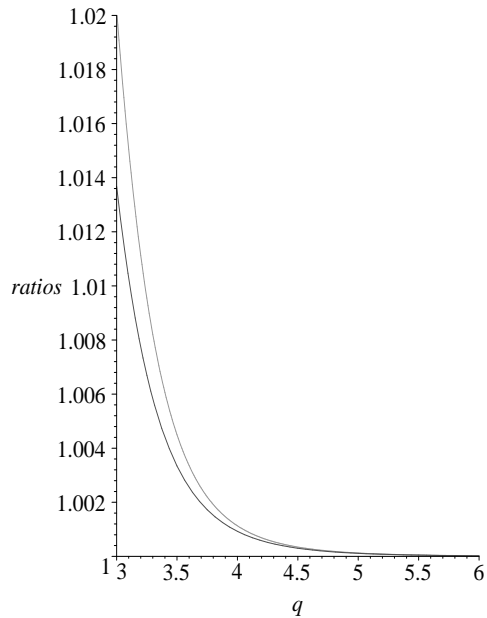}
  \end{center}
\caption{\footnotesize{ Plot of the ratios $R_{sq,q;(2,1)/(1,1)}$ (lower curve)
and $R_{sq,q;(3,1)/(1,1)}$ (upper curve) as functions of $q$ for $3 \le q \le
6$.}}
\label{sqlb_b23k1_ratios}
\end{figure}

\begin{figure}
  \begin{center}
    \includegraphics[height=8cm,width=6cm]{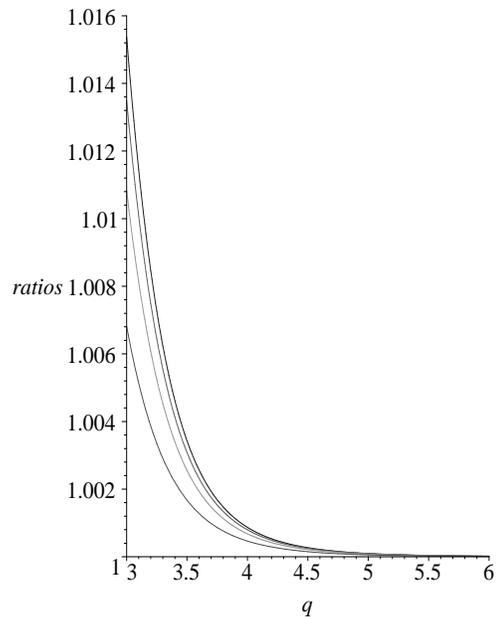}
  \end{center}
\caption{\footnotesize{ Plot of the ratios $R_{sq,q;(1,k)/(1,1)}$ for $k=2$ to
$k=5$ as functions of $q$ for $3 \le q \le 6$. From bottom to top, the curves
    refer to $k=2$, $k=3$, $k=4$, and $k=5$, respectively.}}
\label{sqlb_b1k25_ratios}
\end{figure}

In Fig. \ref{sqlb_b23k1_ratios} we plot the ratios $R_{sq,q;(b,1)/(1,1)}$ for
$b=2$ and $b=3$ as functions of $q$ in the range $3 \le q \le 6$, and in
Fig. \ref{sqlb_b1k25_ratios} we plot the ratios $R_{sq,q;(1,k)/(1,1)}$ for
$k=2$ up to $k=5$, as functions of $q$ in the same range. (Here and below, such
plots entail a continuation of the relevant expressions from integral $q$ to
real $q$.)  These plots illustrate the result that we have proved in general,
that, for a given $q$, $R_{\Lambda,q;(1,k)/(1,1)}$ is an increasing function of
$k$, and also our result that $R_{sq,q;(3,1)/(1,1)} \ge R_{sq,q;(2,1)/(1,1)}$.
(If formally continued below $q=3$ to $q=2$, the curves reach maxima and then
decrease; for example, $R_{sq,q;(2,1)/(1,1)}$ reaches a maximum of 1.06 at $q
\simeq 2.29$ and then decreases to 1 as $q \searrow 2$, while
$R_{sq,q;(1,2)/(1,1)}$ reaches a maximum of 1.03 at $q \simeq 2.29$ and then
decreases to 1 as $q \searrow 2$.) 
As the results in these figures show, our new lower bounds improve most on the
earlier $W(sq,q)_{\ell;1,1}$ in the region of $q \gsim 3$; as $q$ increases
beyond this region, the new bounds approach the earlier one.  This feature will
be evident from the large-$q$ (small-$y$) expansions, since the new bound and
the earlier one coincide in the terms of the small-$y$ expansion up to
$O(y^6)$.  We also find this type of behavior for the new lower bounds that we
have derived for other lattices; that is, the degree of improvement is greatest
for the region of moderate $q$ slightly above $\chi(\Lambda)$.  On a given
lattice $\Lambda$, for larger $q$, our new bounds rapidly approach the earlier
one with $k=1$ and $b=1$; i.e., the ratio $R_{\Lambda,q;(b,k)/(1,1)}$ rapidly
approaches unity.

Combining these results with the results in Table I in \cite{ww} and Table I in
\cite{w3}, it follows that as $q$ increases above the interval of $q=3$ and
$q=4$, these lower bounds approach extremely close to the actual respective
values of $W(sq,q)$. As was evident from these tables in \cite{ww,w3}, in the
range $q \ge 3$, the greatest deviation of the lower bound $W(sq,q)_{\ell;1,1}$
from the actual value of $W(sq,q)$ occurs at $q=3$.  It is thus of interest to
determine how much closer our improved lower bounds are to $W(sq,3)$. From our
general expression for $W(sq,q)_{\ell;1,2}$, we calculate the $q=3$ value
\beq
W(sq,3)_{\ell;1,2}{}\Big |_{q=3} = \frac{\sqrt{5+\sqrt{17}}}{2} = 
1.510223959...
\label{wsqlbound_b1k2_value_q3}
\eeq
so that
\beqs
R_{sq,3;\ell;1,2} &\equiv& \frac{W(sq,3)_{\ell;1,2}{}\Big |_{q=3}}{W(sq,3)} 
\cr\cr
& = & \frac{3\sqrt{3(5+\sqrt{17})}}{16} = 0.9809192...
\label{wsqlbound_bb1k2_value_q3_over_exactvalue}
\eeqs
This ratio and the other ones discussed here are listed in Table 
\ref{rsqlbound_bk_q3_table}. 

% ======================================================================

\section{Triangular Lattice}
\label{tri}

\subsection{$b=1$, $k=1$}

Since $W(tri,3)=1$ is exactly known, we will restrict our 
consideration of lower bounds to the range $q \ge 4$. 
We recall that for $b=1$ and $k=1$, one has the lower bound \cite{ww}-\cite{wn}
$W(tri,q) \ge W(tri,q)_{\ell;1,1}$, where
\beq
W(tri,q)_{\ell;1,1} = \frac{(q-2)^2}{q-1} \ . 
\label{wtrilbound_b1k1_value}
\eeq
As was discussed in \cite{ww}, $q$ increases beyond the lowest values above
$\chi(tri)=3$, this lower bound rapidly approaches the known value of
$W(tri,q)$ (see Table I in \cite{ww}), where the latter was determined by a
numerical evaluation of an integral representation and infinite product
expression \cite{baxter87}.  For example, for $q=5, \ 6, \ 7$, 
$R_{tri,q;1,1}$ is equal to 0.9938, 0.9988, and 0.9996, respectively, and it
increases monotonically with larger $q$.  Since our new lower bounds on
$W(tri,q)$ are more restrictive than (\ref{wtrilbound_b1k1_value}), they are 
therefore even closer to the respective actual values of $W(tri,q)$.

The corresponding lower bound on $\overline W(tri,y)$ is 
$\overline W(tri,y) \ge \overline W(tri,y)_{\ell;1,1}$, where 
\beq
\overline W(tri,y)_{\ell;1,1} = (1-y^2)^2
\label{wytrilbound_b1k1_value}
\eeq
(see Table III in \cite{wn}).

% ========================================================================

\subsection{ $b=2, \ 3$, $k=1$ } 

Here we derive a new lower bound on $W(tri,q)$ using our first generalization
of the CCM method with $b=2$, $k=1$.  For this purpose, we need the chromatic
polynomial of the cyclic strip of the triangular lattice of width $L_y=3$
vertices and arbitrary length, $L_x$.  This was calculated in \cite{t}. The
dominant $\lambda$ in (\ref{pformcyc}) is
\beqs
& & \lambda_{tri,3,0,1} = \frac{1}{2} \bigg [ q^3-7q^2+18q-17 \cr\cr
& + & \Big [ q^6-14q^5+81q^4-250q^3+442q^2-436q+193 \Big ]^{1/2} \ \bigg ] \ . 
\cr\cr
& & 
\label{lamtri3_dom}
\eeqs
Combining this with $\lambda_{tri,2,0,1}=(q-2)^2$, we derive the lower bound 
\beq
W(tri,q) \ge W(tri,q)_{\ell;2,1} \ , 
\label{wtrilbound_b2k1}
\eeq
where
\begin{widetext}
\beq
W(tri,q)_{\ell;2,1} = \frac{\lambda_{tri,3,0,1}}{\lambda_{tri,2,0,1}} 
= \frac{\bigg [ q^3-7q^2+18q-17 
+ \Big [ q^6-14q^5+81q^4-250q^3+442q^2-436q+193 \Big ]^{1/2} \ \bigg ]}
{2(q-2)^2} \ . 
\label{wtrilbound_b2k1_value}
\eeq
The reduced function $\overline W(tri,y)$ is given by Eq. (\ref{wbar}) with
$\Lambda=tri$ and $\Delta=6$.  The corresponding lower bound is
\beq
\overline W(tri,y) \ge \overline W(tri,y)_{\ell;2,1} \ , 
\label{wytrilbound_b2k1}
\eeq
where
\beq
\overline W(tri,y)_{\ell;2,1} = \frac{(1+y)^2\bigg [ 1-4y+7y^2-5y^3 + 
\Big [ 1-8y+26y^2-46y^3+53y^4-42y^5+17y^6 \Big ]^{1/2} \ \bigg ]}{2(1-y)^2} 
\ . 
\label{wytrilbound_b2k1_value}
\eeq
\end{widetext}

Our new lower bound $W(tri,q)_{\ell;2,1}$ is larger than, and hence more
restrictive than the previous lower bound, $W(tri,q)_{\ell;1,1}$. That is, from
the analytic forms (\ref{wytrilbound_b1k1_value}) and
(\ref{wytrilbound_b2k1_value}), we have proved that (for $q \ge 4$)
\beq
R_{tri,q;(2,1)/(1,1)} > 1 \ . 
\label{rtrilbound_b2k1_over_b1k1}
\eeq
This ratio approaches 1 as $q \to \infty$. 

As was evident in Table I in \cite{ww}, the deviation of $W(tri,q)_{\ell;1,1}$
from the actual value of $W(tri,q)$ was greatest for $q=4$.  Hence, it is of
interest to determine how much closer our new lower bound 
$W(tri,q)_{\ell;2,1}$ is to the $W(tri,q)$ for this value, $q=4$. 
A closed-form integral representation has been given for $W(tri,q)$
\cite{baxter87}; in particular, an explicit result is the value for $q=4$:
\beq
W(tri,4) = \frac{3 \Gamma(1/3)^3}{4\pi^2} = \frac{2\pi}{\sqrt{3} \, 
\Gamma(2/3)^3} = 1.460998486...
\label{wtri_q4}
\eeq
where the equivalence follows from the relation \newline
$\Gamma(z)\Gamma(1-z) = \pi/\sin(\pi z)$ 
for the Euler Gamma function. We recall that 
\beq
W(tri,4)_{\ell;1,1} = \frac{4}{3} 
\label{wtrilbound_b1k1_q4}
\eeq
so 
\beq
R_{tri,4;\ell;1,1} = \frac{2\Gamma(2/3)^3}{\sqrt{3} \, \pi} = 0.9126178746...
\label{rtri_b1k1_q4}
\eeq
(see Table I of \cite{ww}). The value of our new lower bound at $q=4$ is
\beq
W(tri,4)_{\ell;2,1} = \frac{7+\sqrt{17}}{8} = 1.3903882...
\label{wtrilbound_b2k1_q4}
\eeq
so 
\beq
R_{tri,4;\ell;2,1} = 
\frac{(7+\sqrt{17} \, ) \sqrt{3} \, \Gamma(2/3)^3}{16\pi} = 
0.951669845...
\label{rtri_b2k1_q4}
\eeq

We have also calculated the lower bound $W(tri,q)_{\ell;b,1}$ for $b=3$ and
evaluated this for $q=4$. For reference, we list the various ratios
$R_{tri,4;\ell;b,k}$ in Table \ref{rtrilbound_bk_q4_table}.  We see that
$W(tri,4)_{\ell;2,1}$ and $W(tri,4)_{\ell;3,1}$ are closer to the exact value
of $W(tri,4)$ than $W(tri,4)_{\ell;1,1}$.

% =====================================================================

\subsection{ $b=1$, $2 \le k \le 5$ }

By the same means as above, we derive 
\beq
W(tri,q) \ge W(tri,q)_{\ell;1,2} \ , 
\label{wtrilbound_b1k2}
\eeq
with 
\beq
W(tri,q)_{\ell;1,2} = \bigg [ \frac{\lambda_{tri,3,0,1}}{q-1} \bigg ]^{1/2}  \
, 
\label{wtrilbound_b1k2_value}
\eeq
where $\lambda_{tri,3,0,1}$ was given in Eq. (\ref{lamtri3_dom}).
Equivalently, 
\beq
\overline W(tri,y) \ge \overline W(tri,y)_{\ell;1,2} \ , 
\label{wytrilbound_b1k2}
\eeq
where
\begin{widetext}
\beq
\overline W(tri,y)_{\ell;1,2} = \frac{1}{\sqrt{2}} \, (1+y)^2 \, 
\bigg [ 1-4y+7y^2-5y^3 + 
\Big [ 1-8y+26y^2-46y^3+53y^4-42y^5+17y^6 \Big ]^{1/2} \ \bigg ]^{1/2} \ . 
\label{wytrilbound_b1k2_value}
\eeq
\end{widetext}

For $b=1$, $k=3$, we need the dominant $\lambda$ in the chromatic polynomial
for the cyclic strip of the triangular lattice of width $L_y=k+1=4$, namely,
$\lambda_{tri,4,0,1}$.  This chromatic polynomial was calculated in \cite{t},
and the dominant $\lambda$ is given as the largest root of the quartic equation
(\ref{triquartic}) in Appendix \ref{higherdegree}.  This is also the dominant
$\lambda$ in the chromatic polynomial of the free strip of the triangular
lattice with width $L_y=4$ and arbitrary length \cite{strip}. We have also
calculated $W(tri,q)_{\ell;1,k}$ for $k=4, \ 5$. 
For reference, we list the various ratios $R_{tri,4;\ell;1,k}$ in 
Table \ref{rtrilbound_bk_q4_table}.

% =======================================================================

\subsection{Plots} 

\begin{figure}
  \begin{center}
    \includegraphics[height=8cm,width=6cm]{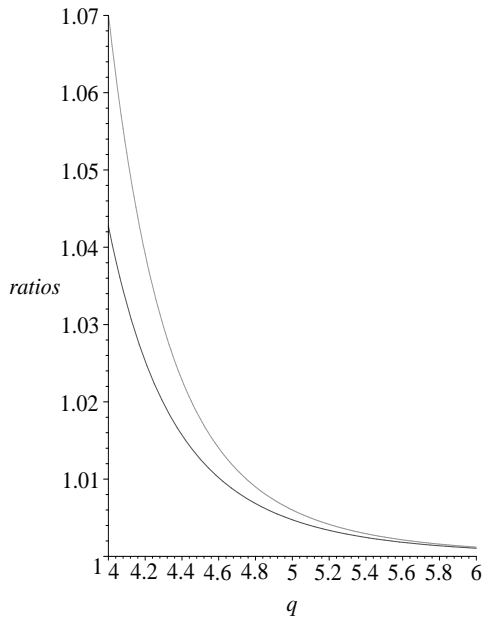}
  \end{center}
\caption{\footnotesize{ Plot of the ratios $R_{tri,q;(2,1)/(1,1)}$ (lower
curve) and $R_{tri,q;(3,1)/(1,1)}$ (upper curve) as functions of $q$ for $4 \le
q \le 6$.}}
\label{trilb_b23k1_ratios}
\end{figure}

\begin{figure}
  \begin{center}
    \includegraphics[height=8cm,width=6cm]{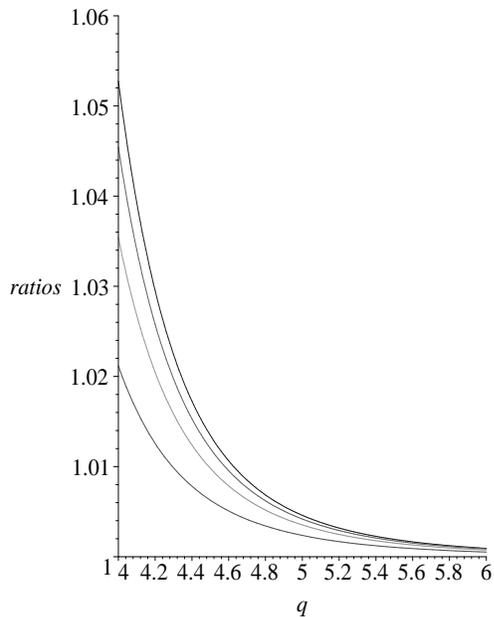}
  \end{center}
\caption{\footnotesize{ Plot of the ratios $R_{tri,q;(1,k)/(1,1)}$ for $k=2$ to
$k=5$ as functions of $q$ for $4 \le q \le 6$. From bottom to top, the curves
    refer to $k=2$, $k=3$, $k=4$, and $k=5$, respectively.}}
\label{trilb_b1k25_ratios}
\end{figure}

In Fig. \ref{trilb_b23k1_ratios} we plot the ratios $R_{tri,q;(b,1)/(1,1)}$ for
$b=2$ and $b=3$ as functions of $q$ in the range $4 \le q \le 6$, and in
Fig. \ref{trilb_b1k25_ratios} we plot the ratios $R_{tri,q;(1,k)/(1,1)}$ for
$k=2$ up to $k=5$, as functions of $q$ in same range. As with the square
lattice, these plots illustrate the result that we have proved in general,
that, for a given $q$, $R_{\Lambda,q;(1,k)/(1,1)}$ is an increasing function of
$k$, and also our result that $R_{tri,q;(3,1)/(1,1)} \ge
R_{tri,q;(2,1)/(1,1)}$.
%

% ======================================================================

\section{Honeycomb Lattice}
\label{hc}

Since $W(hc,2)=1$ is exactly known, we restrict our
consideration of lower bounds for the honeycomb lattice to the range $q \ge 3$.
We recall that for $b=1$ and $k=1$, one has the lower bound $W(hc,q) \ge
W(hc,q)_{\ell;1,1}$, where \cite{w3}
\beq
W(hc,q)_{\ell;1,1} = \frac{(D_6)^{1/2}}{q-1} = 
\frac{(q^4-5q^3+10q^2-10q+5)^{1/2}}{q-1} \ , 
\label{whclbound_b1k1_value}
\eeq
where the general expression for $D_n$ is given in Eq. (\ref{dk}). 
Ref. \cite{w3} noted that as $q$ increases beyond the lowest values above
$\chi(hc)=2$, this lower bound rapidly approaches the actual value of
$W(hc,q)$ (see Table I in \cite{w3}), where the latter was determined by a
Monte-Carlo simulation checked for larger $q$ with a large-$q$ series
approximation. For example, for example, for $q=3, \ 4, \ 5$, 
$R_{hc,q;1,1}$ is equal to 0.99898, 0.99985, and 0.99996, respectively, and it
increases monotonically with larger $q$.  Since our new lower bounds on
$W(hc,q)$ are more restrictive than (\ref{whclbound_b1k1_value}), they are 
therefore even closer to the respective actual values of $W(hc,q)$.

The corresponding lower bound on $\overline W(hc,y)$ is 
$\overline W(hc,y) \ge \overline W(hc,y)_{\ell;1,1}$, where \cite{wn} 
\beq
\overline W(hc,y)_{\ell;1,1} = (1+y^5)^{1/2}
\label{wyhclbound_b1k1}
\eeq
(see Table III in \cite{wn}). 

For the calculation of $W(hc,q)_{\ell;2,1}$, we need the chromatic polynomial
of the cyclic strip of the honeycomb lattice of width $L_y=2b-1=3$ vertices and
arbitrary length, $L_x$, in particular, the dominant $\lambda$.  This
$\lambda_{hc,3,0,1}$ is the largest (real) root of the cubic equation
(\ref{hc_cubic}) in Appendix \ref{higherdegree} \cite{hca}. 
This dominant $\lambda$ is also the input that we need for the calculation of
$W(hc,q)_{\ell;1,2}$, since the latter requires the same chromatic polynomial
of the cyclic strip of the honeycomb lattice of width $L_y=k+1=3$ vertices and
arbitrary length, $L_x$, in particular, the dominant term.  This $\lambda$ is
also the dominant term in the chromatic polynomial of the strip of the
honeycomb lattice of width $L_y=3$ vertices and arbitrary length, with free 
boundary conditions \cite{strip}.

% =====================================================================

\section{$4 \cdot 8^2$ Lattice}
\label{488}

Using the CCM method with $b=1$ and $k=1$, Ref. \cite{w3} derived the lower
bound $W((4 \cdot 8^2),q) \ge W((4 \cdot 8^2),q)_{\ell;1,1}$, where
\beq
W((4 \cdot 8^2)),q)_{\ell;1,1} = \frac{(D_4D_8)^{1/4}}{q-1} \ .
\label{w488lbound_b1k1}
\eeq
Equivalently, $\overline W((4 \cdot 8^2),y) > 
\overline W((4 \cdot 8^2),y)_{\ell;1,1}$, where 
\beq
\overline W((4 \cdot 8^2),y)_{\ell;1,1} = [(1+y^3)(1+y^7)]^{1/4} \ . 
\label{wy488lbound_b1k1}
\eeq

We have obtained the slightly more restrictive 
lower bound $W((4 \cdot 8^2),q) \ge 
W((4 \cdot 8^2),q)_{\ell;1,2}$, where 
\beq
W((4 \cdot 8^2),q)_{\ell;1,2} = \bigg [ \frac{\lambda_{(4 \cdot 8^2),3,0,1}}
{q-1} \bigg ]^{1/3} \ , 
\label{w488lb_b1k2_value}
\eeq
where $\lambda_{(4 \cdot 8^2),3,0,1}$ is the largest (real) root of the cubic
equation (\ref{488_cubic}) in Appendix \ref{higherdegree}.  Correspondingly, 
$\overline W((4 \cdot 8^2),y) \ge 
\overline W((4 \cdot 8^2),y)_{\ell;1,2}$.  We analyze the small-$y$ expansion
of $\overline W((4 \cdot 8^2),y)_{\ell;1,2}$ below. 

% =========================================================================

\section{$3 \cdot 6 \cdot 3 \cdot 6$ (kagom\'e) Lattice}
\label{kag}

In this section we consider the $(3 \cdot 6 \cdot 3 \cdot 6)$ lattice, commonly
called the kagom\'e lattice (which we shall abbreviate as $kag$).  Using the
CCM method with $b=1$ and $k=1$, Ref. \cite{wn} derived the lower bound
$W(kag,q) \ge W(kag,q)_{\ell;1,1}$, where
\beq
W(kag,q)_{\ell;1,1} = \frac{D_3^{2/3}D_6^{1/3}}{q-1} \ .
\label{wkaglbound_b1k1_value}
\eeq
Equivalently, $\overline W(kag,y) > \overline W(kag,y)_{\ell;1,1}$, where
\cite{wn} 
\beq
\overline W(kag,y)_{\ell;1,1} = (1-y^2)^{2/3}(1+y^5)^{1/3} \ . 
\label{wykaglbound_b1k1_value}
\eeq
The zigzag path used in the derivation of this lower bound was
described in detail in Ref. \cite{wn}. 
\begin{figure}
  \begin{center}
    \includegraphics[height=8cm,width=6cm]{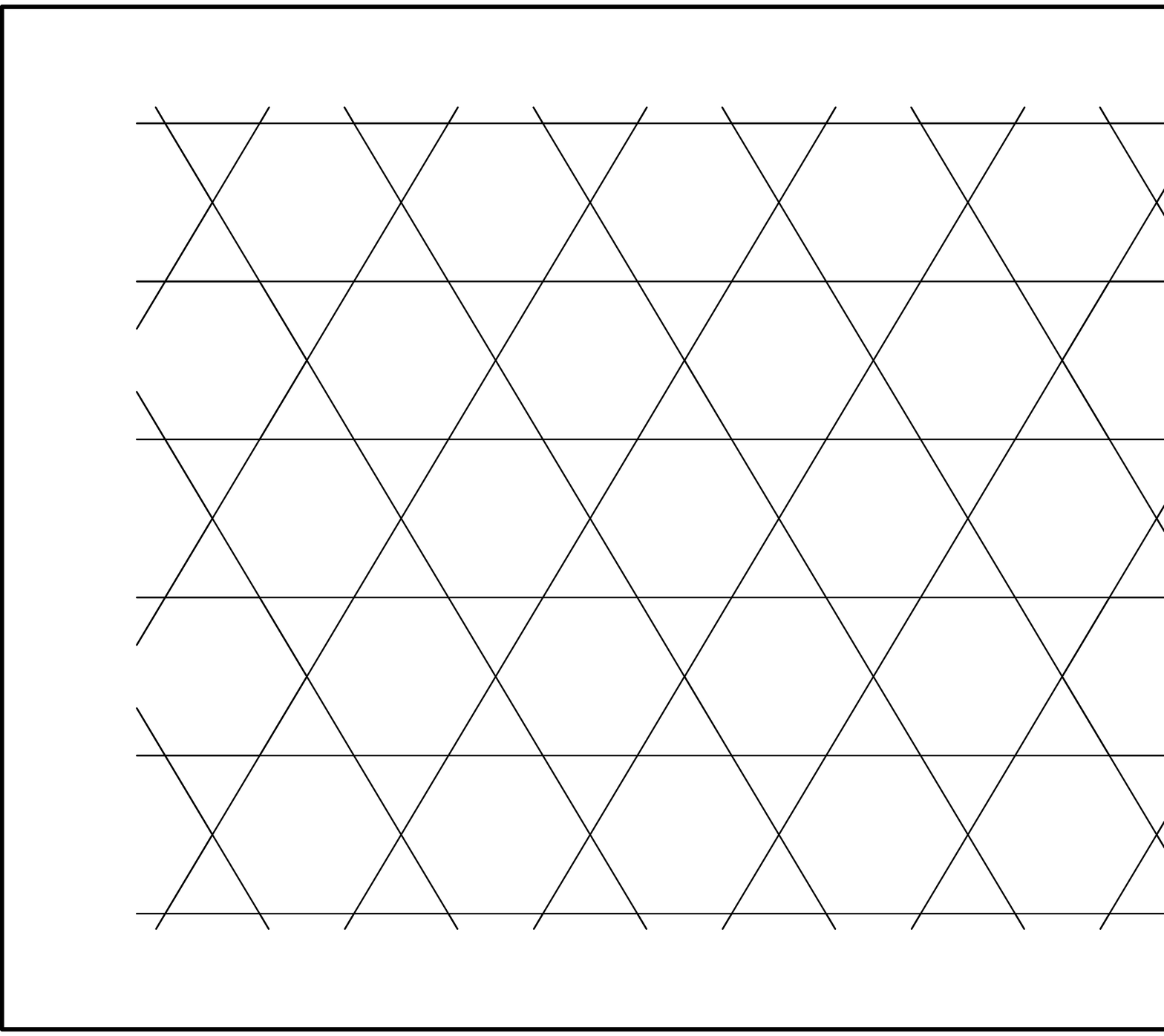}
  \end{center}
\caption{\footnotesize{ Section of the $(3 \cdot 6 \cdot 3 \cdot 6)$ (kagom\'e)
    lattice.}}
\label{fig3636}
\end{figure}
Here, we again take $b=1$ and $k=1$ but use a different type of path. A section
of the kagom\'e lattice is shown in Fig. \ref{fig3636}.  Rather than the zigzag
path used in \cite{wn}, we choose the path to be given horizontal line in
Fig. \ref{fig3636}.  The matrix $T$ then links the proper $q$-coloring of the
vertices on this line, the vertices between this line and, say, the line above
it, and the vertices on this higher-lying horizontal line.  It
turns out that the use of this different path yields a slightly more
restrictive lower bound, which we shall indicate with a prime, namely $W(kag,q)
\ge W(kag,q)_{\ell;1,1}'$, where
\begin{widetext}
\beqs
& & W(kag,q)_{\ell;1,1}' = \cr\cr
& & \Bigg [ \frac{(q-2)\bigg [ q^4-6q^3+14q^2-16q+10+
\Big [q^8-12q^7+64q^6-200q^5+404q^4-548q^3+500q^2-292q+92 \Big ]^{1/2} 
\bigg ]}{2(q-1)^2} \Bigg ]^{1/3} \ .
\cr\cr
& & 
\label{wkaglboundprime_b1k1_value}
\eeqs
Equivalently, we have  $\overline W(kag,y) \ge \overline 
W(kag,y)_{\ell;1,1}'$, where
\beqs
& & \overline W(kag,y)_{\ell;1,1}' = \cr\cr
& & 2^{-1/3} (1+y)\Bigg [ (1-y)\bigg [
1-2y+2y^2-2y^3+3y^4+ \Big [1-4y+8y^2-12y^3+14y^4-16y^5+16y^6-12y^7+9y^8 
\Big ]^{1/2}  \bigg ] \Bigg ]^{1/3} \ . 
\cr\cr
& & 
\label{wykaglboundprime_b1k1_value}
\eeqs
\end{widetext} 
We find that 
\beq
W(kag,q)_{\ell;1,1}' \ge W(kag,q)_{\ell;1,1}
\label{wkagkagrel}
\eeq

The fact that the use of a different path can yield
a more restrictive bound with the same value of $b$ and $k$ was already shown
for the honeycomb lattice in \cite{ww,w3}. Thus, both Ref. \cite{w3} and
Ref. \cite{ww} used the CCM method with $b=1$ and $k=1$, but Ref. \cite{w3}
obtained a more restrictive lower bound for the honeycomb lattice by using a
different path.  The bounds $W(kag,q)_{\ell;1,1}$ and $W(kag,q)_{\ell;1,1}'$
both rapidly approach the actual value of $W(kag,q)$ as $q$ increases beyond
the chromatic number, $\chi(kag)=3$. Below we shall show how the slight 
improvement with the new bound is manifested in the respective small-$y$ 
expansions of $W(kag,q)_{\ell;1,1}$ and $W(kag,q)_{\ell;1,1}'$.
In passing, we note that we have also studied generalizations of the CCM method
for some other Archimedean lattices. 

% ======================================================================

\section{$sq_d$ Lattice}
\label{sqd}

So far, we have considered planar lattices.  The coloring compatibility matrix
method and our generalizations of it, also apply to a subclass of nonplanar
lattices, namely the subclass that can be constructed starting from a planar
lattice and adding edges between vertices on the original planar lattice. An
example of this is the $sq_d$ lattice. As noted above, the $sq_d$ lattice is
formed from the square lattice by adding edges (bonds) connecting the two sets
of diagonal next-nearest-neighbor vertices in each square. Thus, the vertices
and edges in each square form a $K_4$ graph. (Here, the $K_N$ graph is the
graph with $N$ vertices such that each vertex is connected to every other
vertex by one edge.) Although an individual $K_4$ graph is planar, the $sq_d$
lattice is nonplanar.  This lattice has coordination number $\Delta_{sq_d}=8$
and chromatic number $\chi(sq_d)=4$.  Although it is not 4-partite, an analysis
of the way in which the number of proper 4-colorings of the vertices of a
section of the $sq_d$ lattice grows with its area shows that $W(sq_d,4)=1$.

Using the $b=1$, $k=1$ CCM, Ref. \cite{w3} derived the lower bound
$W(sq_d,q) \ge W(sq_d,q)_{\ell;1,1}$, where
\beq
W(sq_d,q)_{\ell;1,1} = \frac{\lambda_{sq_d,2,0,1}}{\lambda_{C,0,1}} = 
\frac{(q-2)(q-3)}{q-1} \ . 
\label{wsqdlbound_b1k1_value}
\eeq
%

% =======================================================================

\subsection{$b=2$, $k=1$} 

For our first generalization, namely $b =2$ and $k=1$, we need the dominant
$\lambda$ for a cyclic strip of the $sq_d$ lattice of width $L_y=3$, which is
\cite{k}
\beqs
& & \lambda_{sq_d,3,0,1} = \frac{(q-3)}{2}\bigg [ q^2-6q+11 \cr\cr
& + & \Big [q^4-12q^3+54q^2-112q+97 \Big ]^{1/2} \ \bigg ] \ .
\label{lamsqd3_dom}
\eeqs
We thus derive the new lower bound $W(sq_d,q) \ge W(sq_d,q)_{\ell;2,1}$, where
$W(sq_d,q)_{\ell;2,1} = \lambda_{sq_d,3,0,1}/\lambda_{sq_s,2,0,1}$, i.e., 
\beqs
& & W(sq_d,q)_{\ell;2,1} = \cr\cr
& &  \frac{q^2-6q+11 
+ \Big [q^4-12q^3+54q^2-112q+97 \Big ]^{1/2} }{2(q-2)} \ . \cr\cr
& & 
\label{wsqdlbound_b2k1_value}
\eeqs
From these explicit analytic results, we find
\beq
R_{sq_d,q;(2,1)/(1,1)} > 1 \ . 
\label{wsqd_b2k1_over_b1k1}
\eeq
That is, our new lower bound $W(sq_d,q)_{\ell,2,1}$ is larger and hence more
restrictive than the one obtained in \cite{w3}. 

The corresponding lower bounds for the reduced $W$ functions are 
$\overline W(sq_d,y) \ge \overline W(sq_d,y)_{\ell;1,1}$, where 
\beqs
\overline W(sq_d,y)_{\ell;1,1} & = & (1-y)(1-2y)(1+y)^3 \cr\cr
                               & = & 1-4y^2-2y^3+3y^4+2y^5 \cr\cr
& & 
\label{wysqd_lbound_b1k1}
\eeqs
and $\overline W(sq_d,y) \ge \overline W(sq_d,y)_{\ell;2,1}$, where 
\begin{widetext}
\beq
\overline W(sq_d,y)_{\ell;2,1} = \frac{(1+y)^3 \bigg [ 1-4y+6y^2 
+ \Big [1-8y+24y^2-36y^3+28y^4 \Big ]^{1/2} \ \bigg ] } {2(1-y)} \ . 
\label{wysqd_lbound_b1k1_value}
\eeq
%

% =======================================================================

\subsection{$b=1$, $k=2$} 

For $b=1$ and $k=2$, we derive the lower bound $W(sq_d,q) \ge
W(sq_d,q)_{\ell;1,2}$, where
\beq
W(sq_d,q)_{\ell;1,2} = \Big [ \frac{\lambda_{sq_d,3,0,1}}{\lambda_{C,0,1}} 
\Big ]^{1/2} = 
\Bigg [ \frac{(q-3)\Big [ q^2-6q+11 + \Big (q^4-12q^3+54q^2-112q+97 
\Big )^{1/2} \Big ] }{2(q-1)} \Bigg ]^{1/2} \ . 
\label{wsqdlbound_b1k2_value}
\eeq
Equivalently, $\overline W(sq_d,y) \ge \overline W(sq_d,y)_{\ell;1,2}$, where
\beq
\overline W(sq_d,y)_{\ell;1,2} = \frac{1}{\sqrt{2}} (1+y)^3 \bigg [ (1-2y)
\Big [1-4y+6y^2+\Big (1-8y+24y^2-36y^3+28y^4 \Big )^{1/2} \Big ] \bigg ]^{1/2}
\ . 
\label{wysqd_lbound_b1k2_value}
\eeq
\end{widetext}

% =====================================================================

\section{Small-$y$ Expansions of New Lower Bounds} 
\label{small_y_series}

\subsection{General}

A lower bound on a function such as $W(\Lambda,q)$ or $\overline W(\Lambda,y)$
plays a role that is different from, and complementary to, that of a Taylor
series expansion, in this case, a small-$y$ expansion.  The lower bound is
valid for any value of $q$ that is physical, but need not, {\it a priori}, be
an accurate approximation to the actual function.  In contrast, the large-$q$
(equivalently, small-$y$) Taylor series expansion is an approximation to the
function itself and, within its radius of convergence, it satisfies the usual
Taylor series convergence properties. Thus, if one truncates this series to a
fixed order of expansion, then it becomes a progressively more accurate
approximate as the expansion variable becomes smaller, and for a fixed value of
the expansion variable, it becomes a more accurate expansion as one includes
more terms.

A lower bound on a function $\overline W(\Lambda,y)$ need not, {\it a priori},
agree with the terms in the small-$y$ Taylor series expansion of this function.
Some explicit examples of this are given in Appendix \ref{rpartite}.
Interestingly, as discussed in \cite{ww}-\cite{wn}, the lower bounds derived
there do agree with these small-$y$ series to a number of orders in $y$ (listed
for Archimedean lattices in Table III and for the duals of Archimedean lattices
in Table IV of Ref. \cite{wn}). 

It is thus clearly of interest to carry out a similar comparison to determine
the extent to which our new lower bounds, which we have shown improve upon
those in \cite{biggs77} and \cite{ww}-\cite{wn}, agree with the respective
small-$y$ expansions to higher order.  We do this in the present section,
showing that our new lower bounds are not only more stringent than the earlier
ones, but also agree with the small-$y$ expansions of $\overline W(\Lambda,y)$
to higher order in $y$ than these earlier lower bounds.

Because $\overline W(\Lambda,y)_{\ell;b,k}$ is a lower bound on 
$\overline W(\Lambda,y)$, one can draw one immediate inference concerning the
comparison of the small-$y$ Taylor series for these two functions, namely that
for a given lattice $\Lambda$, if the small-$y$ Taylor series of 
$\overline W(\Lambda,y)_{\ell;b,k}$ coincides with the 
small-$y$ series for $\overline W(\Lambda,y)$ to order $O(y^{i_c})$, inclusive,
then the difference 
\beq
\overline W(\Lambda,y) - \overline W(\Lambda,y)_{\ell;b,k} = \kappa_{i_c+1} 
y^{i_c+1}, \quad {\rm with} \ \kappa_{i_c+1} > 0 \ .
\label{yseriesdif}
\eeq
Thus, for example, with the $O(y^{i_c+1})$ term in 
$\overline W(\Lambda,y)$ denoted $w_{\Lambda,i_c+1}$
and with the $O(y^{i_c+1})$ term in 
$\overline W(\Lambda,y)_{\ell;b,k}$ denoted $w_{\Lambda;b,k;i_c+1}$, we have
\beq
w_{\Lambda;i_c+1} \ge w_{\Lambda;b,k;i_c+1} \ .
\label{winequality}
\eeq

We discuss a subtlety in this comparison.  One should first show that the
small-$y$ expansion is, in fact, a Taylor series expansion, i.e., that
$\overline W(\Lambda,y)$ is an analytic function at $y=0$ in the complex $y$
plane, or equivalently, that $W_r(\Lambda,q)$ is an analytic function at
$1/q=0$ in the complex plane of the variable $1/q$. In fact, there are families
of $N$-vertex graphs $G_N$ such that $W_r(\{ G \},q)$ is not analytic at
$1/q=0$ \cite{wa23}, where here $\{ G \}$ denotes the formal limit $\lim_{N \to
\infty} G_N$. This is a consequence of the property that the accumulation set
of zeros of the chromatic polynomial $P(G_N,q)$, denoted ${\cal B}$, extends
to infinite $|q|$ in the $q$ plane, or equivalently, to the point $1/q=0$ in
the $1/q$ plane.  (The zeros of $P(G,q)$ are denoted as the chromatic zeros of
$G$.)  Refs. \cite{wa23} constructed and analyzed various families of
graphs for which this is the case. For regular (vertex-transitive) $N$-vertex
graphs $G_{\Lambda,N}$ of a lattice $\Lambda$ with either free or periodic (or
twisted periodic) boundary conditions, the resultant $W_r(\Lambda,q)$ functions
obtained in the $N \to \infty$ limit are analytic at $1/q=0$. This follows
because a necessary condition that ${\cal B}$ extends to infinitely large $|q|$
as $N \to \infty$ is that the chromatic zeros of $G_{\Lambda,N}$ have
magnitudes $|q| \to \infty$ in this limit.  However, a vertex-transitive graph
$G$ has the property that all vertices have the same degree, $\Delta$ and a
chromatic zero of $G$ has a magnitude bounded above as $|q| < 8.4 \Delta$
\cite{sokalbound}. So for the $N \to \infty$ limit of a regular lattice graph
$\Lambda$, $W_r(\Lambda,q)$ is analytic at $1/q=0$ and equivalently, $\overline
W(\Lambda,y)$ is analytic at $y=0$, and the corresponding series expansions in
powers of $1/q$ and powers of $y$ are Taylor series expansions.

% =========================================================================

\subsection{Square Lattice}
\label{sqyseries}

The small-$y$ expansion of $\overline W(sq,y)$ is \cite{kimenting} 
\beqs
& & \overline W(sq,y) = 1 + y^3 + y^7 + 3y^8 + 4y^9 + 3y^{10} \cr\cr
& + & 3y^{11} + O(y^{12}) \ .
\label{wysq_yseries}
\eeqs
This series and several others for regular lattices are known to higher order
than we list; we only display the various series up to the respective orders
that are relevant for the comparison with our lower bounds. As is evident from
Eq. (\ref{wysqlbound_b1k1_value}), the previous lower bound $\overline
W(sq,y)_{\ell;1,1}=1+y^3$ \cite{biggs77} coincides with the small-$y$ series to
$O(y^6)$, inclusive.

We list below the small-$y$ expansions of the various new lower bound functions
$\overline W(sq,y)_{\ell;b,k}$ that we have derived with $b \ge 2$ and $k=1$
and with $b=1$, $k \ge 2$:
\beqs
& & \overline W(sq,y)_{\ell;2,1} = 1 + y^3 + y^7 + 3y^8 + 3y^9 + O(y^{10}) 
\cr\cr
& & 
\label{wysq_b2k1_yseries}
\eeqs
\beqs
& & \overline W(sq,y)_{\ell;1,2} = 1 + y^3 + \frac{1}{2}y^7 
+ \frac{3}{2}y^8 + \frac{3}{2}y^9 + O(y^{10}) \cr\cr
& & 
\label{wysq_b1k2_yseries}
\eeqs
and
\beqs
& & \overline W(sq,y)_{\ell;1,3} = 1 + y^3 + \frac{2}{3}y^7 
+ 2y^8 + \frac{7}{3}y^9 + O(y^{10}) \ . \cr\cr
& & 
\label{wysq_b1k3_yseries}
\eeqs

Comparing the small-$y$ expansion of our new lower bound function $\overline
W(sq,y)_{\ell;b,1}$ with $b=2$, as well as the old lower bound function
$\overline W(sq,y)_{\ell;1,1}$, with the actual small-$y$ series for $\overline
W(sq,y)$ in Eq. (\ref{wysq_yseries}), we can make several observations. First,
the small-$y$ expansions for $\overline W(sq,y)_{\ell;2,1}$ coincides with the
small-$y$ expansion of $\overline W(sq,y)$ to $O(y^8)$, inclusive, which is an
improvement by two orders in powers of $y$ as compared with $\overline
W(sq,y)_{\ell;1,1}$ (see Eq. (\ref{wysqlbound_b1k1_value})). Since increasing
$b$ (with $k$ fixed) improves the accuracy of the lower bound, it follows that
$\overline W(sq,y)_{\ell;b,1}$ will also coincide with the series for
$\overline W(sq,y)$ to at least $O(y^8)$ for $b \ge 3$ as well as for $b=2$.
Moreover, although the respective coefficients of $y^9$ in the series for
$\overline W(sq,y)_{\ell;1,1}$ and $\overline W(sq,y)_{\ell;2,1}$, namely 0 and
3, do not match the coefficient of $y^9$ in the actual small-$y$ expansion of
$\overline W(sq,y)$, which is 4, one can see that as $b$ increases from 1 to 2,
this coefficient of the $y^9$ term increases toward the exact coefficient.

Regarding the matching of terms in the small-$y$ expansions of the $\overline
W(sq,y)_{\ell;b,1}$, as compared with $\overline W(sq,y)_{\ell;1,k}$, that we
have calculated, we find that this matching is better by two orders for the
$\overline W(sq,y)_{\ell;b,1}$ than $\overline W(sq,y)_{\ell;1,k}$.  That is,
for the $k$ values that we have calculated, namely $k=2, \ 3$, the lower
bounds $\overline W(sq,y)_{\ell;1,k}$ match the small-$y$ expansion of
$\overline W(sq,y)$ to order $O(y^6)$, the same order as $\overline
W(sq,y)_{\ell;1,1}$. 

 A related property of our lower bounds for a general lattice $\Lambda$ and, in
particular, for the square lattice, follows as a consequence of the theorem
(\ref{merikoski}) and (\ref{wlb_b1kk_increasing}): with $b=1$, since the lower
bound $\overline W(\Lambda,y)_{\ell;1,k}$ is a monotonically increasing
function of $k$, the degree of matching of coefficients in the small-$y$
expansion for $\overline W(\Lambda,y)$ must improve monotonically as $k$ is
increased.  {\it A priori}, this improvement could be manifested in two ways
(or a combination of the two): (i) as $k$ is increased, coefficients of terms
of higher order in $y$ are exactly matched, or (ii) the coefficient of a given
term of a certain order in $y$ approaches monotonically toward the exact value.
For the present lattice $\Lambda=sq$, we see that, for the $\overline
W(sq,q)_{\ell;1,k}$ that we have calculated, the latter type of behavior, (ii),
occurs.  That is, as we increase $k$ from 1 to 2 to 3, the coefficient of the
$y^7$ term in the small-$y$ series for $\overline W(sq,y)_{\ell;1,k}$ increases
from 0 to 1/2 to 2/3, moving toward the exact value of 1.  This is similar to
the behavior that we observed with the respective coefficients of the $y^9$
term in the small-$y$ expansions of $\overline W(sq,y)_{\ell;b,1}$ as compared
with the exact value.  This type of behavior is in accord with the inequality
(\ref{winequality}).

Regarding the relative ordering of the various lower bounds that we have
obtained, from the small-$y$ expansion, we find, for large $q$, the ordering 
\beqs
& & \overline W(sq,y) > \overline W(sq,y)_{\ell;3,1} > \overline
W(sq,y)_{\ell;2,1} \cr\cr
& > & \overline W(sq,y)_{\ell;1,3} > \overline W(sq,y)_{\ell;1,2} >
 \overline W(sq,y)_{\ell;1,1} \ . \cr\cr
& & 
\label{wysqlbound_ordering}
\eeqs
In fact, we find that this ordering also extends down to the lowest value where
we apply our lower bounds, namely $q=3$.  For bounds on $W(sq,4)$ and 
$W(sq,5)$, see \cite{lm}. 

% =========================================================================

\subsection{Triangular Lattice}
\label{triyseries}

The small-$y$ expansion of $\overline W(tri,y)$ is \cite{kimenting} 
\beqs
\overline W(tri,y) & = & 1 - 2y^2 + y^4 + y^5 + 5y^6 + 16y^7 + 47y^8
\cr\cr
& + & 134y^9 + O(y^{10}) \ .
\label{wtri_yseries}
\eeqs
As is evident from Eq. (\ref{wytrilbound_b1k1_value}), the previous lower bound
$\overline W(tri,y)_{\ell;1,1} = (1-y^2)^2$ \cite{ww,wn}, matches the small-$y$
series to $O(y^4)$, inclusive.  

We list below the small-$y$ expansions of the various new lower bounds
$\overline W(tri,y)_{\ell;b,k}$ that we have derived with $b \ge 2$ and $k=1$,
and with $b=1$, $k \ge 2$: 

\beqs
& & \overline W(tri,y)_{\ell;2,1} = 1 - 2y^2 + y^4 + y^5 + 5y^6 + 14y^7 + 
O(y^8) \cr\cr
& & 
\label{wytri_b2k1_yseries}
\eeqs
\beqs
& & \overline W(tri,y)_{\ell;1,2} = 1 - 2y^2 + y^4 + \frac{1}{2}y^5 
+ \frac{5}{2}y^6 + O(y^7) \cr\cr
& & 
\label{wytri_b1k2_yseries}
\eeqs
and
\beqs
& & \overline W(tri,y)_{\ell;1,3} = 1 - 2y^2 + y^4 + \frac{2}{3}y^5 + 
\frac{10}{3}y^6 + O(y^7) \ . \cr\cr
& & 
\label{wytri_b1k3_yseries}
\eeqs
Comparing these with the small-$y$ series for $\overline W(tri,y)_{\ell;1,1}$,
we find that, among (\ref{wytri_b2k1_yseries})-(\ref{wytri_b1k3_yseries}), the
greatest matching of terms is achieved with (\ref{wytri_b2k1_yseries}), i.e.,
by increasing $b$.  Specifically, the small-$y$ expansion for $\overline
W(tri,y)_{\ell;2,1}$ matches the small-$y$ expansion of $\overline W(tri,y)$ to
$O(y^6)$ inclusive, which is an improvement by two orders in $y$ as compared
with $\overline W(tri,y)_{\ell;1,1}$. This increase by two orders in $y$ is the
same amount of improvement that we found for our lower bound for the square
lattice, $\overline W(sq,y)_{\ell;2,1}$ as compared with $\overline
W(sq,y)_{\ell;1,1}$.

As was true of the lower bounds for the square lattice, the lower bounds 
$\overline W(tri,y)_{\ell;1,k}$ with $k=2$ and $k=3$ coincide with the
small-$y$ series for $\overline W(tri,y)$ to the same order, namely $O(y^4)$,
as $\overline W(tri,y)_{\ell;1,1}$.  However, as $k$ increases from 1 to 2 to
3, the coefficient of the first unmatched term in the respective small-$y$ 
series for $\overline W(tri,y)_{\ell;1,k}$, viz., the $y^5$ term, increases
from 0 to 1/2 to 2/3, moving toward the exact value of 1.  An inequality
that follows from the theorem (\ref{merikoski}) and 
general result (\ref{wlb_b1kk_increasing}), is that with $b=1$,
$\overline W(tri,y)_{\ell;1,k}$ is a monotonically increasing function of $k$.

Concerning the relative ordering of the various lower bounds that we have
obtained, from the small-$y$ expansion, we find, for large $q$, the ordering 
\beqs
& & \overline W(tri,y) > \overline W(tri,y)_{\ell;2,1} > 
\overline W(tri,y)_{\ell;1,3} > \cr\cr
& > & \overline W(tri,y)_{\ell;1,2} > \overline W(tri,y)_{\ell;1,1} \ .
\label{wytrilbound_ordering}
\eeqs
Indeed, we find that this ordering also extends down to the lowest value where
we apply our bounds, namely $q=4$. 

% =========================================================================

\subsection{Honeycomb Lattice}

The small-$y$ expansion of $\overline W(hc,y)$ is \cite{kimenting} 
\beqs
\overline W(hc,y) & = & 1 + \frac{1}{2}y^5 - \frac{1}{2^3}y^{10} + y^{11} 
+ 2y^{12} + \frac{3}{2}y^{13} \cr\cr
& + & y^{14} - \frac{15}{2^4}y^{15} + O(y^{16}) \ .
\label{whc_yseries}
\eeqs
The previous lower bound $\overline W(hc,y)_{\ell;1,1} = (1+y^5)^{1/2}$
\cite{ww}-\cite{wn} has the small-$y$ expansion 
\beqs
& & \overline W(hc,y)_{\ell;1,1} = 1 + \frac{1}{2}y^5 - \frac{1}{2^3}y^{10} 
+ \frac{1}{2^4}y^{15} + O(y^{20}) \ . 
\cr\cr
& & 
\label{whclbound_b1k1_yseries}
\eeqs
Thus, as was noted in \cite{ww}-\cite{wn}, this small-$y$ expansion coincides
with the small-$y$ expansion of $\overline W(hc,y)$ to the quite high order
$O(y^{10})$.  

We list below the small-$y$ expansions of the various new lower bound functions
$\overline W(hc,y)_{\ell;b,k}$ that we have derived with $b \ge 2$ and $k=1$
and with $b=1$, $k \ge 2$: 
\beq
\overline W(hc,y)_{\ell;2,1} = 
1 + \frac{1}{2}y^5 - \frac{1}{2^3}y^{10} + y^{11}
+ 2y^{12} + y^{13} + O(y^{15})
\label{wyhc_b2k1_yseries}
\eeq
and
\beq
\overline W(hc,y)_{\ell;1,2} = 1 + \frac{1}{2}y^5 - \frac{1}{2^3}y^{10} 
+ \frac{1}{2}y^{11} + y^{12} + O(y^{13}) \ . 
\label{wyhc_b1k2_yseries}
\eeq
As with the square and triangular lattices, we find that among
(\ref{wyhc_b2k1_yseries})-(\ref{wyhc_b1k2_yseries}), the greatest matching of
terms is achieved with (\ref{wyhc_b2k1_yseries}), i.e., by increasing $b$.
Specifically, the small-$y$ expansion for $\overline W(hc,y)_{\ell;2,1}$
matches the small-$y$ expansion of $\overline W(hc,y)$ to $O(y^{12})$
inclusive, which is an improvement by two orders in $y$ as compared with
$\overline W(hc,y)_{\ell;1,1}$. 

The theorem (\ref{merikoski}) and corollary (\ref{wlb_b1kk_increasing}) imply
that $W(hc,q)_{\ell;1,2} > W(hc,q)_{\ell;1,1}$, and this
inequality is reflected in the degree of matching of the small-$y$ expansions
for the corresponding functions $\overline W(hc,y)_{\ell;1,2}$ and $\overline
W(hc,y)_{\ell;1,1}$. Although $\overline W(hc,y)_{\ell;1,2}$ does not 
increase the order of matching, as compared with $\overline
W(hc,y)_{\ell;1,1}$, it begins the process of building up a nonzero 
coefficient for a $y^{11}$ term, which was zero in the expansion of 
$\overline W(hc,y)_{\ell;1,1}$. Specifically, the small-$y$ expansion of 
$\overline W(hc,y)_{\ell;1,2}$ contains a $y^{11}$ term with coefficient 1/2,
building toward the exact coefficient, 1, of $y^{11}$ in (\ref{whc_yseries}).

% =========================================================================

\subsection{$4 \cdot 8^2$ Lattice}

We next consider a (bipartite) heteropolygonal Archimedean lattice, namely the
$(4 \cdot 8^2)$ lattice. The small-$y$ expansion of $\overline W((4 \cdot
8^2),y)$ is \cite{w3,wn}
\beqs
\overline W((4 \cdot 8^2),y) & = & 1 + \frac{1}{4}y^3 - \frac{3}{2^5} y^6 
+ \frac{1}{4} y^7 + \frac{7}{2^7} y^9 \cr\cr
& + & \frac{1}{2^4}y^{10} - \frac{77}{2^{11}}y^{12} + O(y^{13}) \ . 
\label{w488_yseries}
\eeqs
The small-$y$ expansion of the lower bound obtained in \cite{ww,wn}, 
$\overline W((4 \cdot 8^2),y)_{\ell;1,1}$, is
\beqs
& & \overline W((4 \cdot 8^2),y)_{\ell;1,1} = 1 + \frac{1}{4}y^3 
- \frac{3}{2^5}y^6 + \frac{1}{4}y^7 + \frac{7}{2^7}y^9 \cr\cr
& + & \frac{1}{2^4}y^{10} - \frac{77}{2^{11}}y^{12} - \frac{3}{2^7} y^{13} 
- \frac{3}{2^5}y^{14} + \frac{231}{2^{13}}y^{15} + O(y^{16}) \ . \cr\cr
& & 
\label{wylb488_b1k1_yseries}
\eeqs
As was noted in \cite{w3,wn}, this coincides with the small-$y$ expansion of 
$\overline W((4 \cdot 8^2),y)$ to the quite high order $O(y^{12})$. 

We list below the small-$y$ expansions of the various new lower bound functions
$\overline W((4\cdot 8^2),y)_{\ell;b,k}$ that we have derived with $b \ge 2$ 
and $k=1$ and with $b=1$, $k \ge 2$: 
\beqs
& & \overline W((4 \cdot 8^2),y)_{\ell;2,1} = 1 + \frac{1}{4}y^3 
- \frac{3}{2^5}y^6 + \frac{1}{4}y^7 \cr\cr
& + & \frac{7}{2^7}y^9 + \frac{1}{2^4}y^{10} - 
\frac{77}{2^{11}}y^{12} + \frac{189}{2^7}y^{13} + O(y^{14}) 
\cr\cr
& & 
\label{wy488_b2k1_yseries}
\eeqs
and
\beqs
& & \overline W((4 \cdot 8^2),y)_{\ell;1,2} =  1 + \frac{1}{4}y^3 
- \frac{3}{2^5}y^6 + \frac{1}{4}y^7 \cr\cr
& + & \frac{7}{2^7}y^9 
+ \frac{1}{2^4}y^{10} - \frac{77}{2^{11}}y^{12} + \frac{93}{2^7}y^{13}
+ \frac{45}{32}y^{14} + O(y^{15}) \ . 
\cr\cr
& & 
\label{wy488_b1k2_yseries}
\eeqs
Evidently, the small-$y$ series expansions of 
$\overline W((4 \cdot 8^2),y)_{\ell;2,1}$ and 
$\overline W((4 \cdot 8^2),y)_{\ell;1,2}$ match the small-$y$ expansion of 
$\overline W((4 \cdot 8^2),y)$ to at least the same order as 
$\overline W((4 \cdot 8^2),y)_{\ell;1,1}$. Further, we observe that for
small-$y$, 
\beq
\overline W((4 \cdot 8^2),y)_{\ell;2,1} > 
\overline W((4 \cdot 8^2),y)_{\ell;1,2} > 
\overline W((4 \cdot 8^2),y)_{\ell;1,1}
\label{wylb488_ordering}
\eeq
%

% =========================================================================

\subsection{$3 \cdot 6 \cdot 3 \cdot 6$ (kagom\'e) Lattice}

The small-$y$ expansion of $\overline W(kag,y)$ is \cite{wn} 
\beqs
& & \overline W(kag,y) = 1 - \frac{2}{3} y^2 - \frac{1}{3^2}y^4+\frac{1}{3}y^5
\cr\cr
& & - \frac{4}{3^4}y^6 - \frac{2}{3^2} y^7 - \frac{7}{3^5}y^8 + O(y^9) \ .
\label{wkag_yseries}
\eeqs
As was discussed in \cite{wn}, the small-$y$ expansion of the $b=1$, $k=1$
lower bound $\overline W(kag,y)_{\ell;1,1}$ derived there (listed above as
Eq. (\ref{wkaglbound_b1k1_value})) coincides to $O(y^8)$ with the small-$y$
series for the actual quantity $\overline W(kag,y)$.  Explicitly, 
\beqs
& & \overline W(kag,y)_{\ell;1,1} = 1 - \frac{2}{3} y^2 - \frac{1}{3^2}y^4 + 
\frac{1}{3}y^5
\cr\cr
& & - \frac{4}{3^4}y^6 - \frac{2}{3^2} y^7 - \frac{7}{3^5}y^8 
- \frac{1}{3^3}y^9  - \frac{95}{3^6}y^{10} \cr\cr
& & + O(y^{11}) \ . 
\label{wkaglbound_yseries}
\eeqs
Our new bound has the small-$y$ expansion 
\beqs
& & \overline W(kag,y)_{\ell;1,1}' = 1 - \frac{2}{3} y^2 - \frac{1}{3^2}y^4 + 
\frac{1}{3}y^5
\cr\cr
& & - \frac{4}{3^4}y^6 - \frac{2}{3^2} y^7 - \frac{7}{3^5}y^8 
+ \frac{8}{3^3}y^9  + \frac{634}{3^6} y^{10} \cr\cr
& & + O(y^{11}) \ . 
\label{wkaglboundprime_yseries}
\eeqs
Thus, 
\beq
\overline W(kag,y)_{\ell;1,1}' - \overline W(kag,y)_{\ell;1,1} = \frac{1}{3^2}
y^9 + O(y^{10}) \ . 
\label{wykagdif}
\eeq
One could derive similar lower bounds for other Archimedean lattices not
considered here, e.g., the $3 \cdot 12 \cdot 12$ lattice \cite{wn,tsai312}.

% =========================================================================

\subsection{$sq_d$ Lattice}

Since the lower bound $\overline W(sq_d,y)_{\ell;1,1}$ derived in \cite{w3} and
given above in Eq. (\ref{wysqd_lbound_b1k1}) is a polynomial, it is identical
to its small-$y$ Taylor series expansion.

Expanding $\overline W(sq_d,y)_{\ell;2,1}$, we find 
\beqs
& & \overline W(sq_d,y)_{\ell;2,1} = 1-4y^2-y^3+6y^4+5y^5+O(y^6) \ . \cr\cr
& & 
\label{wysqd_lbound_b1k1_yseries}
\eeqs
Similarly, 
\beqs
& & \overline W(sq_d,y)_{\ell;1,2} = 1 - 4y^2 - \frac{3}{2}y^3 + 
\frac{9}{2}y^4 + \frac{7}{2}y^5 + O(y^6) \cr\cr
& & 
\label{wysqd_lbound_b1k2_yseries}
\eeqs
\beqs
& & \overline W(sq_d,y)_{\ell;1,3} = 1 - 4y^2 - \frac{4}{3}y^3 + 
5 y^4 + \frac{13}{3}y^5 + O(y^6)  \ . \cr\cr
& & 
\label{wysqd_lbound_b1k3_yseries}
\eeqs
From these expansions we find, for large $q$, the ordering
\beq
\overline W(sq_d,y) > \overline W(sq_d,y)_{\ell;2,1} > 
\overline W(sq_d,y)_{\ell;1,2} > \overline W(sq_d,y)_{\ell;1,1}
\label{wysq_ordering}
\eeq
This is the same ordering that we found for the other lattices. 

% =========================================================================

\section{Conclusions}
\label{conclusions} 

Nonzero ground-state entropy per site, $S_0$, and the associated ground-state
degeneracy per site, $W = e^{S_0/k_B}$, are of fundamental importance in
statistical mechanics.  In this paper we have presented generalized methods for
deriving lower bounds on the ground-state degeneracy per site, $W(\Lambda,q)$,
of the $q$-state Potts antiferromagnet on several different lattices $\Lambda$.
Our first generalization is to consider a coloring compatibility matrix that
relates a strip of width $b \ge 2$ vertices to an adjacent strip of the same
width.  Our second generalization is to consider a coloring compatibility
matrix that acts $k \ge 2$ times in relating a path on $\Lambda$ to an adjacent
parallel path.  We have applied these generalizations to obtain new lower
bounds on $W(\Lambda,q)$, denoted $W(\Lambda,q)_{\ell;b,k}$. In this notation,
the lower bounds previously derived in \cite{ww}-\cite{biggs77} have $b=1$ and
$k=1$.  One of the interesting properties of these bounds
$W(\Lambda,q)_{\ell;1,1}$ obtained in \cite{ww}-\cite{biggs77} was that as $q$
increases beyond $\chi(\Lambda)$ they rapidly approach quite close to the
actual respective values of $W(\Lambda,q)$. We have shown that our new lower
bounds are slightly more restrictive than these previous lower bounds, and
consequently are even closer to the actual values $W(\Lambda,q)$.  We have
demonstrated how this is manifested in the matching to higher-order terms with
the large-$q$ (small-$y$) Taylor series expansions for the corresponding
functions $\overline W(\Lambda,y)$ for the various lattices that we have 
considered.

% ======================================================================

\begin{acknowledgments}

This research was partly supported by the Taiwan Ministry of Science and 
Technology grant MOST 103-2918-I-006-016 (S.-C.C.) and by 
the U.S. National Science Foundation grant No. NSF-PHY-13-16617 (R.S.). 

\end{acknowledgments}

% =====================================================================

\begin{appendix}

\section{$W(\Lambda_{rp.},q)$ at $q=\chi(\Lambda)$}
\label{rpartite}

We mention here a subtlety that results from the noncommutativity in the limits
(\ref{wnoncomm}).  An $r$-partite $rp.$ graph with $N$ vertices, $G_{rp.,N}$
has chromatic number $\chi(G_{rp.,N})=r$.  One equivalent definition of an
$r$-partite graph is that its chromatic polynomial, evaluated at $q=r$,
satisfies
\beq
P(G_{rp.,N}, r) = r! 
\label{pgrpartite}
\eeq
The square and honeycomb lattices are bipartite (as are the $(4 \cdot 6 \cdot
12)$, and $(4 \cdot 8 \cdot 8)$ lattices, among Archimedean lattices), while
the triangular lattice is tripartite (for others Archimedean lattices and their
planar duals, see, e.g., Tables I and II in \cite{wn}). It follows that, with
the $D_{Nq}$ definition for $W(\Lambda,q)$, namely setting $q=r$ and then
taking the $N \to \infty$ limit in Eq. (\ref{w}), one has
\beq
W(\Lambda_{rp.},r)=1 \ . 
\label{wrpartite}
\eeq
As discussed in \cite{w}, because of the noncommutativity (\ref{wnoncomm}), if
instead of setting $q=r$, evaluating $P(G_{rp.},r)$, and then taking the $N
\to \infty$, one first takes $N \to \infty$ with $q$ in the vicinity of $r$,
and then performs the limit $q \to r$, one can, in general, get a different
result for $W(\Lambda,q)$. Indeed, this is the case for many lattice strips of
regular lattices of a fixed width $L_y$, an arbitrary length, $L_x$, and
various transverse and longitudinal boundary conditions
\cite{w,wcy,s4,t}.  The coloring problem on a given lattice $\Lambda$ is of 
interest for $q \ge \chi(\Lambda)$, since this is the minimum (integer) value
of $q$ for which one can carry out a proper $q$-coloring of the vertices of
$\Lambda$. In a number of cases, $\chi(\Lambda) <
q_c(\Lambda)$. If one considers $W(\Lambda,q)$ for $q < q_c(\Lambda)$, then
one must deal with the generic noncommutativity in the limits (\ref{wnoncomm})
\cite{w}.  Here we always use the order $D_{Nq}$, i.e., we fix $q$ to a given
value and then take $N \to \infty$.  Actually, in view of the results 
(\ref{pgrpartite}) and (\ref{wrpartite}), for the square and honeycomb
lattices, $W(sq,2)=W(hc,2)=1$, and for the triangular lattice, 
$W(tri,3)=1$.  Since our new lower bounds are intended for practical use
and since one already knows (with the $D_{Nq}$ definition) the values of 
$W(sq,2)$, $W(hc,2)$, and $W(tri,3)$ exactly, we may restrict our analysis to
the application of our new bounds in the range $q \ge 3$ for the square and
honeycomb lattices and to the range $q \ge 4$ for the triangular lattice. 

  For reference, we recall an elementary lower bound on $P(G,q)$ and hence
on $\lim_{N \to \infty}P(G,q)^{1/N}$, where $G$ is an $N$-vertex graph. If $G$
is bipartite ($bp.$) then one can assign a color to all of the vertices of the
even subgraph in any of $q$ ways and then one can assign one of the remaining
$q-1$ colors to each of the vertices on the odd subgraph independently, so
$P(G_{bp.},q) \ge q(q-1)^{N/2}$.  Hence, for a bipartite lattice, denoting
$\Lambda_{bp.}$ as the $N \to \infty$ limit of $G_{bp.}$, one has
$W(\Lambda_{bp.},q) \ge (q-1)^{1/2}$.  Both of these lower bounds are realized
as equalities only in the case $q=2$.  More generally, if $G_{rp.}$ is an
$r$-partite graph and $\Lambda_{rp.} = \lim_{N \to \infty} \Lambda_{rp.,N}$, 
then
\beq
P(G_{rp.},q) \ge \Big [ \prod_{s=0}^{r-2} (q-s) \Big ]
[q-(r-1)]^{N/r} 
\label{plb_rpartite}
\eeq
and hence
\beq
W(\Lambda_{rp.},q) \ge \ [q-(r-1)]^{1/r} \ . 
\label{wlb_rpartite}
\eeq
Thus, for example, one has
the elementary lower bounds $W(sq,q) \ge (q-1)^{1/2}$ and $W(tri,q) \ge
(q-2)^{1/3}$, etc.

For $q > r$ on $\Lambda_{r p.}$, the lower bound
(\ref{wlb_rpartite}) is less stringent than the ones derived in
\cite{wn}-\cite{biggs77} and here via coloring matrix methods. Indeed, these
lower bounds illustrate the fact noted in the text, namely that, {\it a
priori}, a lower bound need not agree with terms in the large-$q$ expansion 
of $W_r(\Lambda,q)$ or the equivalent small-$y$ expansion of $\overline
W(\Lambda,y)$. For example, for the square and honeycomb lattices, the 
$r=2$ special cases of (\ref{wlb_rpartite}) read, for $q \ge 2$,
\beq
W(sq,q) \ge (q-1)^{1/2}
\label{wsqlb_2partite}
\eeq
and
\beq
W(hc,q) \ge (q-1)^{1/2} \ . 
\label{whclb_2partite}
\eeq
Since $W(\Lambda,q) \sim q$ for large $q$, these lower bounds becomes 
progressively worse (i.e., farther from the actual value) as $q$ increases 
above 2. The corresponding lower bounds in terms of $\overline W(sq,y)$ and
$\overline W(hc,y)$ are
\beq
\overline W(sq,y) \ge (1+y)\sqrt{y}
\label{wysqlb2partite}
\eeq
and
\beq
\overline W(hc,y) \ge \sqrt{y(1+y)} \ .
\label{wyhclb2partite}
\eeq
Rather than matching any terms in the respective small-$y$ expansions
(\ref{wysq_yseries}) and (\ref{whc_yseries}), the right-hand sides of these
lower bounds vanish for small $y$.  Similarly, since the triangular lattice is
tripartite, the $r=3$ special case of (\ref{wlb_rpartite}) yields the lower
bound, for $q \ge 3$, 
\beq
W(tri,q) \ge (q-2)^{1/3} \ . 
\label{wtrilbound_3partite}
\eeq
In terms of $\overline W(tri,y)$, this is 
\beq
\overline W(tri,y) \ge y^{2/3}(1-y)^{1/3}(1+y)^2 \ . 
\label{wytrilb3partite}
\eeq
Again, for small $y$, this vanishes rather than matching any of the terms of
the small-$y$ expansion (\ref{wtri_yseries}). Thus, as noted, a lower
bound need not match any of the terms in the small-$y$ expansion.  This
emphasizes how impressive the new lower bounds are in their matching of these
terms in the small-$y$ expansions for the various lattices to high order.

% ========================================================================

% Appendix B
\section{Lower Bounds $W(\Lambda,q)_{\ell;1,1}$ and 
$\overline W(\Lambda,y)_{\ell;1,1}$ for Archimedean Lattices}
\label{wnbounds} 

A number of general results were proved in Ref. \cite{wn} concerning lower
bounds which, in the notation of this paper, are $W(\Lambda,q)_{\ell;1,1}$ and
$\overline W(\Lambda,y)_{\ell;1,1}$.  These results in \cite{wn} applied for
all eleven Archimedean lattices. We have given the notation for an Archimedean
lattice in the text.  These lower bounds from \cite{wn} are relevant here
because we compare our new lower bounds $W(\Lambda,q)_{\ell;b,k}$ and the
corresponding lower bounds $\overline W(\Lambda,y)_{\ell;b,k}$ with $b \ge 2$
and/or $k \ge 2$ to these earlier ones with $b=k=1$. (Ref. \cite{wn} also gave
lower bounds for the planar duals of the Archimedean lattices; we do not list
these here but instead refer the reader to \cite{wn}.)

The chromatic polynomial of a circuit graph is $P(C_n,q) = (q-1)^n +
(q-1)(-1)^n$.  Since this chromatic polynomial has $q(q-1)$ as a factor, we can
write it as $P(C_n,q) = q(q-1)D_n(q)$, where
\beq
D_n(q) = \frac{P(C_n,q)}{q(q-1)} = \sum_{s=0}^{n-2}(-1)^s {{n-1}\choose {s}} \,
q^{n-2-s} \ . 
\label{dk}
\eeq

Ref. \cite{wn} proved the following general lower bounds for an Archimedean
lattice, $\Lambda=(\prod_i p_i^{a_i})$ (where we add the subscripts $1,1$ to
indicate $b=1$ and $k=1$ to match our current notation for
$W(\Lambda,q)_{\ell;b,k}$): 
\beq
W \Bigl ( (\prod_i p_i^{a_i}),q \Bigr ) \ge 
W \Bigl ( (\prod_i p_i^{a_i}),q \Bigr )_{\ell;1,1} \ , 
\label{wlbound_b1k1_wn}
\eeq
where
\beq
W \Bigl ( (\prod_i p_i^{a_i}),q \Bigr )_{\ell;1,1} =
\frac{\prod_i D_{p_i}(q)^{\nu_{p_i}}}{q-1} \ , 
\label{wlbound_b1k1_wn_value}
\eeq
Here, the $\{i\}$ in the product label the set of $p_i$-gons involved in
$\Lambda$ and $\nu_{p}$ (with $p=p_i$ here) was defined in Eq. (2.10) of
\cite{wn}.

This lower bound takes a somewhat simpler form in
terms of the related function $\overline W(\Lambda,y)_\ell$, namely,
\beq
\overline W \Bigl ( (\prod_i p_i^{a_i}),y \Bigr ) \ge 
\overline W \Bigl ( (\prod_i p_i^{a_i}),y \Bigr )_{\ell;1,1} \ , 
\label{wylbound_b1k1_wn}
\eeq
where

\beq
\overline W \Bigl ( (\prod_i p_i^{a_i}),y \Bigr )_{\ell,1,1} =
\prod_i \Bigl [ 1+(-1)^{p_i}y^{p_i-1} \Bigl ]^{\nu_{p_i}}
\label{wylbound_b1k1_wn_value}
\eeq
A summary of these for Archimedean lattices is given in Table IV of \cite{wn}.

% =====================================================================

\section{Higher-Degree Algebraic Equations for Certain
  $\lambda_{\Lambda,L_y,0,1}$}
\label{higherdegree} 

In this appendix we list some algebraic equations of degree higher than 2 that
are used in the text. The cubic equation whose largest (real) root is 
$\lambda_{sq,4,0,1}$, used for our lower bound $W(sq,q)_{\ell;1,3}$, is
\begin{widetext}
\beqs
& & \lambda^3 - (q^4-7q^3+23q^2-41q+33)\, \lambda^2 
+ (2q^6-23q^5+116q^4-329q^3+553q^2-517q+207)\, \lambda \cr\cr
& - & q^8+16q^7-112q^6+449q^5-1130q^4+1829q^3-1858q^2+1084q-279 = 0 \ . 
\label{eqsq4}
\eeqs
The quartic equation whose largest (real) root is $\lambda_{tri,4,0,1}$, used
for our lower bound $W(tri,q)_{\ell;1,3}$, is 
\beqs
& & \lambda^4 - (q^4-10q^3+42q^2-88q+76)\, \lambda^3 + 
(q-2)(q-3)^2(3q^3-22q^2+60q-60)\, \lambda^2 \cr\cr
& - & (q-2)^2(q-3)^3(3q^3-21q^2+51q-43)\, \lambda + (q-2)^6(q-3)^4 = 0 \ . 
\label{triquartic}
\eeqs
The cubic equation whose largest (real) root is $\lambda_{hc,3,0,1}$ used in 
our lower bound $W(hc,q)_{\ell;2,1}$ is 
\beqs
& & \lambda^3 - (q^6-8q^5+28q^4-56q^3+71q^2-58q+26)\, \lambda^2 \cr\cr
& + & (q-1)^2(q^6-10q^5+43q^4-102q^3+144q^2-120q+49)\, \lambda 
- (q-1)^4(q-2)^2 = 0 \ .
\label{hc_cubic}
\eeqs
The cubic equation whose largest (real) root is 
$\lambda_{(4 \cdot 8^2),3,0,1}$, used in our bound $W((4 \cdot
8^2),q)_{\ell;1,2}$ is 
\beqs
& & \lambda^3 -(q^{12}-16q^{11}+120q^{10}-558q^9+1794q^8-4212q^7+7437q^6
-10018q^5+10324q^4 \cr\cr
& & -8064q^3+4648q^2-1854q+414)\, \lambda^2 \cr\cr
& & +(q-1)^4(q^{12}-20q^{11}+188q^{10}-1094q^9+4375q^8-12640q^7+27033q^6
-43164q^5 \cr\cr 
& & +51235q^4 -44380q^3+26931q^2-10462q+2017)\, \lambda \cr\cr
& & - (q-1)^8(q-2)^2(q-3)^2(q^2-3q+3)^2=0 \ . 
\label{488_cubic}
\eeqs
\end{widetext}

\end{appendix}

% =======================================================================

% =======================================================================

\newpage

\begin{table}
\caption{\footnotesize{Values of $R_{sq,q;\ell;b,k}$ for $q=3$ and some 
illustrative values of $b$ and $k$.}}
\begin{center}
\begin{tabular}{|c|c|c|c|}
\hline\hline
$b$ & $k$ & $W(sq,3)_{\ell;b,k}$ & $R_{sq,3;\ell;b,k}$ 
\\ \hline
1  &  1  &  1.500000  &  0.974279  \\
2  &  1  &  1.520518  &  0.987605  \\
3  &  1  &  1.530340  &  0.993985  \\
\hline
1  &  2  &  1.510224  &  0.980919  \\
1  &  3  &  1.5162645 &  0.984843  \\
1  &  4  &  1.520249  &  0.987430  \\
1  &  5  &  1.523073  &  0.989265  \\
\hline\hline
\end{tabular}
\end{center}
\label{rsqlbound_bk_q3_table}
\end{table}
%

% =====================================================================

\begin{table}
\caption{\footnotesize{Values of $R_{tri,q;\ell;b,k}$ for $q=4$ and some 
illustrative values of $b$ and $k$.}}
\begin{center}
\begin{tabular}{|c|c|c|c|}
\hline\hline
$b$ & $k$ & $W(tri,4)_{\ell;b,k}$ & $R_{tri,4;\ell;b,k}$ 
\\ \hline
1  &  1  &  1.333333  &  0.912618  \\
2  &  1  &  1.390388  &  0.951670  \\
3  &  1  &  1.427052  &  0.976765  \\
\hline
1  &  2  &  1.361562  &  0.931939  \\
1  &  3  &  1.380569  &  0.944949  \\
1  &  4  &  1.393923  &  0.954089  \\
1  &  5  &  1.403672  &  0.960762  \\
\hline\hline
\end{tabular}
\end{center}
\label{rtrilbound_bk_q4_table}
\end{table}
%

% =======================================================================


\begin{thebibliography}{99}

% 1 
\bibitem{i35}
W. F. Giauque and J. W. Stout, J. Am. Chem. Soc. {\bf 58},
1144 (1936). Here, $R=N_{Avog.}k_B = 1.99$ cal/(K-mole). 

% 2
\bibitem{paulingbook}
L. Pauling, J. Am. Chem. Soc. {\bf 57}, 2680 (1935);
L. Pauling, {\it The Nature of the Chemical Bond}
(Cornell Univ. Press, Ithaca, 1960), p. 466.

% 3
\bibitem{berg07}
B. A. Berg, C. Muguruma, and Y. Okamoto, Phys. Rev. B {\bf 75}, 092202 (2007).

% 4 
\bibitem{wurev}
F. Y. Wu, Rev. Mod. Phys. {\bf 54}, 235 (1982). 

% 5
\bibitem{ww}
R. Shrock and S.-H. Tsai, Phys. Rev. E {\bf 55}, 6791 (1997).

% 6
\bibitem{w3}
R. Shrock and S.-H. Tsai, Phys. Rev. E {\bf 56}, 2733 (1997). 

% 7
\bibitem{wn}
R. Shrock and S.-H. Tsai, Phys. Rev. E {\bf 56}, 4111 (1997). 

% 8
\bibitem{biggs77}
N. L. Biggs, Bull. London Math. Soc. {\bf 9}, 54 (1977).

% 9 
\bibitem{pf}
See, e.g.,
P. Lancaster and M. Tismenetsky, {\it The Theory of Matrices,
with Applications} (New York, Academic Press, 1985); H. Minc,
{\it Nonnegative Matrices} (New York, Wiley, 1988).

% 10 
\bibitem{london}
D. London, Duke Math. J. {\bf 33}, 511 (1966).

% 11
\bibitem{gsbook}
Gr\"{u}nbaum, B. and Shephard, G. 1989 {\it Tilings and
Patterns: an Introduction} (Freeman, New York, 1989).

% 11
\bibitem{w}
R. Shrock and S.-H. Tsai, Phys. Rev. {\bf E55}, 5165 (1997). 

% 12
\bibitem{cf}
S.-C. Chang and R. Shrock, Physica A {\bf 296}, 131 (2001).

% 13
\bibitem{wcy}
R. Shrock and S.-H. Tsai, Phys. Rev. E {\bf 60}, 3512 (1999);
R. Shrock and S.-H. Tsai, Physica A {\bf 275}, 429-449 (2000).

% 14
\bibitem{pg}
R. Shrock and S.-H. Tsai, J. Phys. A Letts. {\bf 32}, L195 (1999).

% 15
\bibitem{a}
R. Shrock, Phys. Lett. A {\bf 261}, 57 (1999); 
Physica A {\bf 283}, 388 (2000); Discrete Math. {\bf 231}, 421 (2001).

% 16
\bibitem{k}
S.-C. Chang and R. Shrock, Phys. Rev. E {\bf 62}, 4650 (2000).

% 17
% on the trace and sum elements of a matrix 
\bibitem{merikoski84}
J. K. Merikoski, Linear Alg. and Applic. {\bf 60}, 177 (1984). 

% 18
\bibitem{strip}
M. Ro\v{c}ek, R. Shrock, and S.-H. Tsai), Physica A {\bf 252}, 505
(1998).


% 19
\bibitem{kimenting}
%
D. Kim and I. G. Enting, J. Combin. Theory, B {\bf 26}, 327 (1979).  
Note that the function that Kim and Enting denoted 
$\overline W$ for the honeycomb lattice is implicitly defined
per 2-cell (hexagon), not per site, and hence is equal to 
$\overline W(hc,y)^2$. 

% 20
\bibitem{wsqlbound_b2k1_analytic}
%
As it must be, the lower bound $W(sq,q)_{\ell;2,1}$ is a positive real
analytic function of $q$ in the relevant range of $q$, namely,
$q \ge 2$.  We comment on its analytic structure.  The poles from zeros in the
denominator occur at $q=(1/2)(3 \pm \sqrt{2} \, i)$ and the branch-point
singularities in the square root in the numerator occur where the 
factor $(q^2-5q+7)$ vanishes, at $q=(1/2)(5 \pm \sqrt{3} \, i)$, and at the
two pairs of complex-conjugate zeros of the quartic factor, which are
$q=0.58657 \pm 1.14006i$ and $q=1.91343 \pm 1.09797i$, to the indicated
floating-point accuracy.  Similar comments apply for other explicit analytic
expressions for lower bounds given in the text. 

% 21
\bibitem{lenard} A. Lenard, unpublished, as cited in E. H. Lieb,
Phys. Rev. {\bf 162}, 162 (1967).

% 22
\bibitem{s4}
S.-C. Chang and R. Shrock, Physica A {\bf 290}, 402 (2001).

% 23
\bibitem{s5}
S.-C. Chang and R. Shrock, Physica A {\bf 316}, 335 (2002).

% 24
\bibitem{t}
S.-C. Chang and R. Shrock, Ann. Phys. {\bf 290}, 124 (2001).

% 25
\bibitem{baxter87}
R. J. Baxter, J. Phys. A {\bf 20}, 5231 (1987). 

% 26
\bibitem{hca}
S.-C. Chang and R. Shrock, Physica A {\bf 296}, 183 (2001);
J. Stat. Phys. {\bf 130}, 1011 (2008). 

% 27
\bibitem{wa23}
R. Shrock and S.-H. Tsai, Phys. Rev. {\bf E56}, 3935 (1997); 
J. Phys. A {\bf 31}, 9641 (1998); Physica A {\bf 265}, 186 (1999).

% 28
\bibitem{sokalbound}
A. Sokal, Combin. Probab. Comput. {\bf 10}, 41 (2000). 

% 29
\bibitem{lm}
P. Lundow and K. Markstr\"om, Lond. Math. Soc. {\bf 11}, 1 (2008). 

% 30
\bibitem{tsai312}
S.-H. Tsai, Phys. Rev. E {\bf 57}, 2686 (1998). 

\end{thebibliography}
\end{document}